# IN-SEASON PREDICTION OF BATTING AVERAGES: A FIELD TEST OF EMPIRICAL BAYES AND BAYES METHODOLOGIES[1]

By Lawrence D. Brown

*University of Pennsylvania*

Batting average is one of the principle performance measures for an individual baseball player. It is natural to statistically model this as a binomial-variable proportion, with a given (observed) number of qualifying attempts (called "at-bats"), an observed number of successes ("hits") distributed according to the binomial distribution, and with a true (but unknown) value of $p_i$ that represents the player's latent ability. This is a common data structure in many statistical applications; and so the methodological study here has implications for such a range of applications.

We look at batting records for each Major League player over the course of a single season (2005). The primary focus is on using only the batting records from an earlier part of the season (e.g., the first 3 months) in order to estimate the batter's latent ability, $p_i$, and consequently, also to predict their batting-average performance for the remainder of the season. Since we are using a season that has already concluded, we can then validate our estimation performance by comparing the estimated values to the actual values for the remainder of the season.

The prediction methods to be investigated are motivated from empirical Bayes and hierarchical Bayes interpretations. A newly proposed nonparametric empirical Bayes procedure performs particularly well in the basic analysis of the full data set, though less well with analyses involving more homogeneous subsets of the data. In those more homogeneous situations better performance is obtained from appropriate versions of more familiar methods. In all situations the poorest performing choice is the naïve predictor which directly uses the current average to predict the future average.

One feature of all the statistical methodologies here is the preliminary use of a new form of variance stabilizing transformation in order to transform the binomial data problem into a somewhat more familiar structure involving (approximately) Normal random variables

Received September 2007; revised September 2007.
[1]Supported in part by NSF Grant DMS-07-07033.

with known variances. This transformation technique is also used in the construction of a new empirical validation test of the binomial model assumption that is the conceptual basis for all our analyses.

## 1. Introduction.

*Overview.* Batting average is one of the principle performance measures for an individual baseball player. It is the percentage of successful attempts, "Hits," as a proportion of the total number of qualifying attempts, "At-Bats." In symbols, the batting average of the $i$th player may be written as $BA_i = H_i/AB_i$. This situation, with Hits as a number of successes within a qualifying number of attempts, makes it natural to statistically model each player's batting average as a binomial variable outcome, with a given value of $AB_i$ and a true (but unknown) value of $p_i$ that represents the player's latent ability.

As one outcome of our analysis we will demonstrate that this model is a useful and reasonably accurate representation of the situation for Major League players over periods of a month or longer within a given baseball season. (The season is approximately 6 months long.)

We will look at batting records [Brown (2008)] for each Major League player over the course of a single season (2005). We use the batting records from an earlier part of the season (e.g., the first 3 months) in order to estimate the batter's latent ability, $p_i$, and consequently, to predict their BA performance for the remainder of the season. Since we are using a season that has already concluded, we can then validate the performance of our estimator by comparing the predicted values to the actual values for the remainder of the season.

*Dual focus.* Our study has a dual focus. One focus is to develop improved tools for these predictions, along with relative measures of the attainable accuracy. Better mid-season predictions of player's batting averages should enable better strategic performance for managers and players for the remainder of the season. [Of course, other criteria may be equally important, or even more so, as suggested in Albert and Bennett (2001), Lewis (2004), Stern (2005), etc. Some of these other criteria—e.g., slugging percentage or on-base percentage—may be measurable and predictable in a manner analogous to batting average, and so our current experience with batting average may help provide useful tools for evaluation via these other criteria.]

A second focus is to gain experience with the estimation methods themselves as valuable statistical techniques for a much wider range of situations. Some of the methods to be suggested derive from empirical Bayes and hierarchical Bayes interpretations. Although the general ideas behind these techniques have been understood for many decades, some of these methods have only been refined relatively recently in a manner that promises to more accurately fit data such as that at hand.

One feature of all of our statistical methodologies is the preliminary use of a particular form of variance stabilizing transformation in order to transform the binomial data problem into a somewhat more familiar structure involving (approximately) Normal random variables with known variances. This transformation technique is also useful in validating the binomial model assumption that is the conceptual basis for all our analyses. In Section 2 we present empirical evidence about the properties of this transformation and these help provide justification for its use.

*Efron and Morris.* Efron and Morris (1975, 1977) (referred to as E&M in the sequel) presented an analysis that is closely related in spirit and content to ours, but is more restricted in scope and in range of methodology. They used averages from the first 45 at-bats of a sample of 18 players in 1970 in order to predict their batting average for the remainder of the season. Their analysis documented the advantages of using the James–Stein shrinkage in such a situation. [See James and Stein (1961) for the original proposal of this technique.] E&M used this analysis to illustrate the mechanics of their estimator and its interpretation as an empirical Bayes methodology.

In common with the methods we use later, E&M also used a variance stabilizing transformation as a preliminary step in their analysis, but not quite the same one as we propose. Our first stage of data contains some batters with many fewer at-bats, and others with many more. (We only require that a batter have more than 10 first stage at-bats to be included in our analysis.) In such a case the distinction between the various forms of variance stabilization becomes more noticeable.

Because E&M used the same number of initial at-bats (45) for every player in their sample, their transformed data was automatically (approximately) homoscedastic. We do not impose such a restriction, and our transformed data is heteroscedastic with (approximately) known variances. The James–Stein formula adopted in E&M is especially suited to homoscedastic data. We develop alternate formulas that are more appropriate for heteroscedastic data and ordinary squared error. Two of these can be considered direct generalizations within the empirical Bayes framework of the formulas used by E&M. Another is a familiar hierarchical Bayes proposal that is also related to the empirical Bayes structure. The final one is a new type of implementation of Robbins' (1951, 1956) original nonparametric empirical Bayes idea.

E&M also consider a different data context involving toxoplasmosis data, in addition to their baseball data. This setting involves heteroscedastic data. In this context they do propose and implement a shrinkage methodology as a generalization of the James–Stein formula for homoscedastic data. However, unlike the baseball setting, the context of this example does not provide the opportunity to validate the performance of the procedure by comparing predictions with future data (or with the truth). For our batting average data, we include an implementation of a slight variant of the E&M heteroscedastic method. [This is the method referred to in Section 3 and afterward as EB(ML).] We find that it has satisfactory performance in some settings, as in our Table 3, and somewhat less satisfactory performance in others, as in our Table 2. We also propose an explanation for this difference in behavior in relation to the robustness of this procedure relative to an independence assumption involved in its motivation.

*Ground rules.* The present study involves two rather special perspectives in order to restrict considerations to a manageable set of questions. In keeping with the baseball theme of this article, we refer to these as our ground rules.

The first major limitation is that we look *only* at results from the 2005 Major League season. Within this season, we look only at the total numbers of at-bats and hits for each player for each month of the season, and in some parts of our analysis we separate players into pitchers, and all others ("nonpitchers"). It is quite likely that bringing into consideration each player's performance in earlier seasons in addition to their early season performance in 2005 could provide improved predictions of batting performance. But it would also bring an additional range of statistical issues (such as whether batters maintain consistent average levels of ability in successive seasons). There are also many other possible statistical predictors of later season batting average that might be investigated, in addition to the directly obvious values of at-bats and hits and playing position in terms of pitcher/nonpitcher. We hope that our careful treatment of results within our ground rules can assist with further studies concerning prediction of batting average or other performance characteristics.

One may suppose that players—on average—maintain a moderately high level of consistency in performance over successive seasons. If so, performance from prior season(s) could be successfully incorporated to usefully improve the predictive performance of the methods we describe. On the other hand, for working within one given season, our results suggest that later season batting average is an inherently difficult quantity to predict on the basis only of earlier season performance; and hence, it seems somewhat uncertain that other, more peripheral, statistical measures taken only from the earlier part of the season can provide significant overall advantage in addition to knowledge of the player's earlier season record of at-bats and hits, and their position in terms of pitcher/nonpitcher. Possibly, division into other player categories (such as designated hitter, shortstop, etc.) could be useful. This was suggested to us by S. Jensen (private communication) based on his own analyses.

A second guideline for our study is that we concentrate on the issue of estimating the latent ability of each player, with equal emphasis on all players who bat more than a very minimal number of times. This is the kind of goal that would be suitable in a situation where it was desired to predict the batting average of each player in the league, or of each player on a team's roster, with equal emphasis on all players. A contrasting goal would be to predict the batting average of players after weighting each player by their number of at-bats. As we mention in Section 5, and briefly study there, such a goal might favor the use of a slightly different suite of statistical techniques.

As usual, the 2005 Major League regular season ran about 6 months (from April 3 to October 2). It can conveniently be divided into one month segments, beginning with April. The last segment consists of games played in September, plus a few played in very early October. We do not include batting records from the playoffs and World Series. For our purposes, it is convenient to refer to the period in the first three months, April through June, as the "first half" of the season, and the remaining three months (through October 2) as the "second half." (Baseball observers often think of the period up to the All Star break as the first half of the season and the remaining period as the second half. The All Star break in 2005 occurred on July 11–13. We did not split our season in this fashion, but have a checked that doing so would not have a noticeable effect on the main qualitative conclusions of our study.)

*Major questions.* We address several fairly specific questions related to prediction of batting averages:

Q1. Does the player's batting performance during the first half of the season provide a useful basis for predicting his performance during the remainder of the season?

Q2. If so, how can the prediction best be carried out? In particular, is the player's batting average for the first half by itself a useful predictor of his performance for the second half? If not, what is better?

Q3. Is it useful for such predictions to separate categories of batters? The most obvious separation is into two groups—pitchers and all others. Strong batting performance, including a high batting average, is not a priority for pitchers, but is a priority for all other players. So we will investigate whether it is useful, given the player's first half batting average, to perform predictions for the second half batting average of pitchers separately from the prediction for other players.

Q4. What are the answers to the previous questions if one tries to use the player's performance for the first month to predict their performance for the remainder of the season? What if one tries to use the first five months to predict the final month's performance?

Q5. We have already noted an additional question that can be addressed from our data. That is whether the batting performance of individual batters over months of the season can be satisfactorily modeled as independent binomial variables with a constant (but latent) success probability. This is related to the question of whether individual batter's exhibit streakiness of performance. For our purposes, we are interested in whether such streakiness exists, over periods of several weeks or months. If such streakiness exists, then it would be additionally difficult to predict a batter's later season performance on the basis of earlier season performance.

It is unclear whether batting performance exhibits streakiness, and if so, then to what degree. See Albright (1993) and Albert and Bennett (2001), and references therein. Even if some degree of streakiness is present over the range of one or a few games, for all or most batters, this single game streakiness might be supposed to disappear over the course of many games as the effect of different pitchers and other conditions evens out in a random fashion. We will investigate whether this is the case in the data for 2005.

Q6. We have noted the dual focus of our study on the answers to questions such as the above and on the statistical methods that should be used to successfully answer such questions. Hence, we articulate this as a final issue to be addressed in the course of our investigations.

*Answers in brief.* The latter part of this paper provides detailed numerical and graphical answers to the above questions, along with discussion and supporting statistical motivation of the methods used to answer them. However, it is possible to qualitatively summarize the main elements of the answers without going into the detail which follows. Here are brief answers to the major questions, as discussed in the remaining sections of the paper.

A1. The player's first half batting provides useful information as the basis for predictions of second half performance. However, it must be employed in an appropriate manner. The next answer discusses elements of this.

A2. The simplest prediction method that uses the first half batting average is to use that average as the prediction for the second half average. We later refer to this as the "naïve method." This is *not* an *effective* prediction method. In terms of overall accuracy, as described later, it performs worse than simply ignoring individual performances and using the overall mean of batting averages (0.240) as the predictor for all players (sample size, $\mathcal{P} = 567$). [If this batting average figure seems low, remember that we are predicting the average of all players with at least 11 first half at-bats. Hence, this sample contains many pitchers—see A3. The corresponding first-half mean for all nonpitchers ($\mathcal{P} = 499$) is 0.255.]

Table 1
*Mean batting average by pitcher/nonpitcher and half of season*

|  | First half | Second half |
|---|---|---|
| Nonpitchers | 0.255 | 0.252 |
| Pitchers | 0.153 | 0.145 |
| All | 0.240 | 0.237 |

In spite of the above, there are effective ways to use first-half batting performance in order to predict second-half average. These methods involve empirical Bayes or hierarchical Bayes motivations. They have the feature of generally "shrinking" each batter's first-half performance in the general direction of the overall mean. In our situation, the amount of shrinkage and the precise focus of the shrinkage depend on the number of first-half at-bats—the averages of players with fewer first-half at-bats undergo more shrinkage than those with more first-half at-bats. The best of these shrinkage estimators clearly produces better predictions in general than does the use of the overall average, however, it still leaves much more variability that cannot be accounted for by the prediction. See Table 2, where the best method reduces the total sum of squared prediction error, relative to the overall average, by about a 40% decrease in total squared estimation error.

Not all the shrinkage estimators work nearly this well. The worst of the shrinkage proposals turns out to be a minor variant of the method proposed for heteroscedastic data by E&M, and we will later suggest a reason for the poor performance of this shrinkage estimator.

A3. Pitchers and nonpitchers have very different overall batting performance. Table 1 gives the overall mean of the averages for pitchers and for nonpitchers for each half of the season. (In each case, the samples are restricted to those batters with at least 11 at-bats for the respective half of the season.)

It is clear from Table 1 that it is desirable to separate the two types of batters. Indeed, predictions by first half averages separately for pitchers and nonpitchers are much more accurate than a prediction using a single mean of first half averages. Again, see Section 5 for description of the extent of advantage in separately considering these two subgroups.

The empirical Bayes procedures can also be employed within the respective groups of nonpitchers and pitchers. They automatically shrink strongly toward the respective group means, and the better of them have comparable performance to the group mean itself in terms of sum of squared prediction error. See Table 3 for a summary of results.

One of our estimators does particularly well when applied to the full sample of players, and the reasons for this are of interest. This estimator is a nonparametric empirical Bayes estimator constructed according to ideas in Brown and Greenshtein (2007). In essence, this estimator automatically detects (to some degree) that the players with very low first month batting averages should be considered in a different category from the others, and hence, does not strongly shrink the predictions for those batters toward the overall mean. It thus automatically, if imperfectly, mimics the behavior of estimators that separately estimate the ability of players in each of the pitcher or nonpitcher subgroups according to the subgroup means. This estimator and its pattern of behavior will be discussed later, along with that of all our other shrinkage estimators.

A4. One month's records provide much less information about a batter's ability than does a three month record. For this reason, the best among our prediction methods is to just use the overall mean within these subgroups as the prediction within each of the two subgroups. Some of the alternate methods make very similar predictions and have similar performance. The naïve prediction that uses the first month's average to predict later performance is especially poor.

The situation in which one uses records from the first 5 months in order to predict the last month has a somewhat different character. The difference is most noticeable within the group of nonpitchers. Within this group the naïve prediction does almost as well as does the mean within that group; and some of the empirical Bayes estimators perform noticeably better.

A5. Our analyses will show it is reasonably accurate to assume that the monthly totals of hits for each batter are independent binomial variables, with a value of $p$ that depends only on the batter. (The batter's value of $p$ does not depend on the month or on the number of at-bats the batter has within that month, so long as it exceeds our minimum threshold.) The binomial model that underlies our choice of tools for estimation and prediction thus seems well justified. The somewhat limited success of our empirical Bayes tools is primarily an inherent feature of the statistical situation, rather than a flaw in the statistical modeling supporting these tools. A more focused examination of quantitative features inherent in the binomial model makes clear why this is the case.

A6. The preceding answers mention a few features of our statistical methodology. The remainder of the paper discusses the methodology in greater breadth and detail.

Section 2 of the paper discusses variance stabilization for binomial variables. This variance stabilization and normalization is a key building stone for all our analyses. The transformation presented in this section is a variant of the standard methodology, and use of this variant is motivated by the discussion in this section.

Sections 3 and 4 describe further aspects of our estimation and prediction methodology. Section 3 establishes the basic statistical structure for the estimation and validation of the estimators. Section 4 gives definitions and motivation for each of the estimators to be investigated.

Sections 5 and 6 give validation results that describe how well the estimators perform on the baseball data. Section 5 contains a series of results involving estimation based on data from the first three months of the six month season. Section 6 discusses results for estimation based on either the first month or on the first five months of the season.

Section 7 contains a test of the basic assumption that each batter's monthly performance is a binomial random variable with a (latent) value of $p$ that is constant throughout the season. The results of this test confirm the viability of this assumption as a basis for the analyses of Sections 5 and 6.

The Appendix applies this same type of goodness-of-fit test to each batter's performance over shorter ten-day spans. The analyses of Sections 5 and 6 do not require validity of the binomial assumption over such shorter spans of time. However, the issue is of independent interest since it is related to whether batting performance varies over successive relatively short periods. In this case the test procedure identifies a subset of batters whose performance strongly suggests "streakiness" in the sense that their latent ability differs for different ten-day segments of the season.

## 2. Methodology, part I; variance stabilization.

We are concerned with records that tabulate the number of hits and number of at-bats for each of a sample of players over a given period of time. For a given player indexed by $i$, let $H_i$ denote the number of hits and let $N_i$ denote the number of at-bats during the given time period of one or more months. (In Section 1 this was denoted by AB, instead of $N$, but a two-letter symbol is awkward in mathematical displays to follow.) The time period in question should be clear from the text and context of the discussion. Where it is necessary to consider multiple time periods, such as the two halves of the season, we will insert additional subscripts, and use symbols such as $H_{ji}$ or $N_{ji}$ to denote the observed number of hits and at-bats for player $i$ within period $j$. We assume that each $H_i$ is a binomial random variable with an unobserved parameter $p_i$ corresponding to the player's hitting ability. Thus, for data involving $P$ players over two halves of the season, we write

$$(2.1) \qquad H_{ji} \sim Bin(N_{ji}, p_i), \qquad j = 1, 2, i = 1, \ldots, \mathcal{P}_j, \text{indep.}$$

Baseball hitting performance is commonly summarized as a proportion, $R$, called the batting ave**R**age, for which we will use subscripts corresponding to those of its components. (In Section 1 we used the commonsense symbol BA for this.) Thus,

$$R_{ji} = \frac{H_{ji}}{N_{ji}}.$$

Binomial proportions are nearly normal with mean $p$, but with a variance that depends on the unknown value of $p$. For our purposes, it is much more convenient to deal with nearly normal variables having a variance that depends only on the observed value of $N$. The standard variance-stabilizing transformation, $T = \arcsin\sqrt{H/N}$, achieves this goal moderately well. Its lineage includes foundational papers by Bartlett (1936, 1947) and important extensions by Anscombe (1948), as well as Freeman and Tukey (1950) and Mosteller and Youtz (1961), to which we will refer later. It has been used in various statistical contexts, including its use for analyzing baseball batting data in E&M. For purposes like the one at hand, it is preferable to use the transformation

$$(2.2) \qquad\qquad X = \arcsin\sqrt{\frac{H + 1/4}{N + 1/2}}.$$

We will reserve the symbol, $X$, to represent such a variable and, where convenient, will use subscripts corresponding to those for $H$ and $N$.

To understand the advantages of the definition (2.2), consider a somewhat broader family of variance-stabilizing transformations,

$$(2.3) \qquad\qquad Y^{(c)} = \arcsin\sqrt{\frac{H + c}{N + 2c}} \ni H \sim Bin(N, p).$$

Each of these transformations has the variance-stabilizing property

$$(2.4) \qquad\qquad \mathrm{Var}(Y^{(c)}) = \frac{1}{4N} + O\left(\frac{1}{N^2}\right).$$

Anscombe (1948) shows that the choice $c = 3/8$ yields the stronger property, $\mathrm{Var}(Y^{(3/8)}) = (4N + 2)^{-1} + O(N^{-3})$. We will argue that this stronger property is less valuable for our purposes than is the property in (2.6) below.

It can be easily computed that, for each $p : 0 < p < 1$,

$$(2.5) \qquad E(Y^{(c)}) = \arcsin\sqrt{p} + \frac{1 - 2p}{2N\sqrt{p(1 - p)}}(c - 1/4) + O\left(\frac{1}{N^2}\right).$$

Hence, the choice $c = 1/4$ yields

$$E(Y^{(1/4)}) = \arcsin\sqrt{p} + O\left(\frac{1}{N^2}\right).$$

Consequently,

$$(2.6) \qquad \sin^2[E(Y^{(c)})] = p + O\left(\frac{1}{N^2}\right) \Leftrightarrow c = \frac{1}{4}.$$

The transformation with $c = 1/4$ thus gives the best asymptotic control over the bias among all the transformations of the form $Y^{(c)}$. Better control of bias is important to the adequacy of our transformation methods, and is more important than having slightly better control over the variance, as could be yielded by the choice $c = 3/8$. [The transformation proposed by Freeman and Tukey (1950) and studied further by Mosteller and Youtz (1961) is very similar to $Y^{(1/2)}$, and so does not perform as well for our purposes as our preferred choice $Y^{(1/4)}$.]

Results for realistic sample sizes are more pertinent in practice than are asymptotic properties alone. Here, too, the transformation with $c = 1/4$ provides excellent performance both in terms of bias and variance. Figure 1 displays the un-transformed bias of three competing transformations—the traditional one ($c = 0$), the mean-matching one ($c = 1/4$) and Anscombe's transformation ($c = 3/8$). This is defined as

$$(2.7) \qquad Bias = \sin^2(E_p(Y^{(c)})) - p.$$

These plots show that the mean-matching choice ($c = 1/4$) nearly eliminates the bias for $N = 10$ (and even smaller) so long as $0.1 < p < 0.9$. The other transformations require larger $N$ and/or $p$ nearer $1/2$ in order to perform as well. [Plots of $E_p(Y^{(c)}) - \arcsin \sqrt{p}$ exhibit qualitatively similar behavior to those in Figure 1, but we felt plots of (2.7) were slightly easier to interpret. They are also more directly relevant to some of the uses of such transformations, as, e.g., in Brown et al. (2007).] Thus, in later contexts where bias correction is important and control of variance is only a secondary concern, we will restrict attention to batters having more than 10 at-bats.

Figure 2 displays the variance of $Y^{(c)}$ after normalizing by the nominal (asymptotic) value. Thus, the displayed curves are for

$$(2.8) \quad \text{Var ratio}^* = \begin{cases} \mathrm{Var}_p(Y^{(c)})/(\frac{1}{4}N), & \text{for } c = 0,\ 1/4, \\ \mathrm{Var}_p(Y^{(3/8)})/(1/(4N+2)), & \text{for } c = 3/8. \end{cases}$$

Note that in this respect (as well as in terms of bias) $c = 1/4$ and $c = 3/8$ perform much better than does the traditional value $c = 0$. Above about $N = 12$ and $p = 0.150$ both transformations do reasonably well in terms of variance. Nearly all baseball batters can be expected to have $p \geq 0.150$, with the possible exception of some pitchers. In Section 7 where variance stabilization (in addition to low bias) is especially important, we will require for inclusion in our analysis that $N \geq 12$.

It is also of some importance that the transformed variables, $Y$, have approximately a normal distribution, in addition to having very nearly the desired, nominal mean and variance. Even for $N = 12$ and $0.75 \geq p \geq 0.25$ (roughly) the variables for $c = 1/4$ and $c = 3/8$ appear reasonably well approximated by their nominal normal distribution in spite of their very discrete nature.

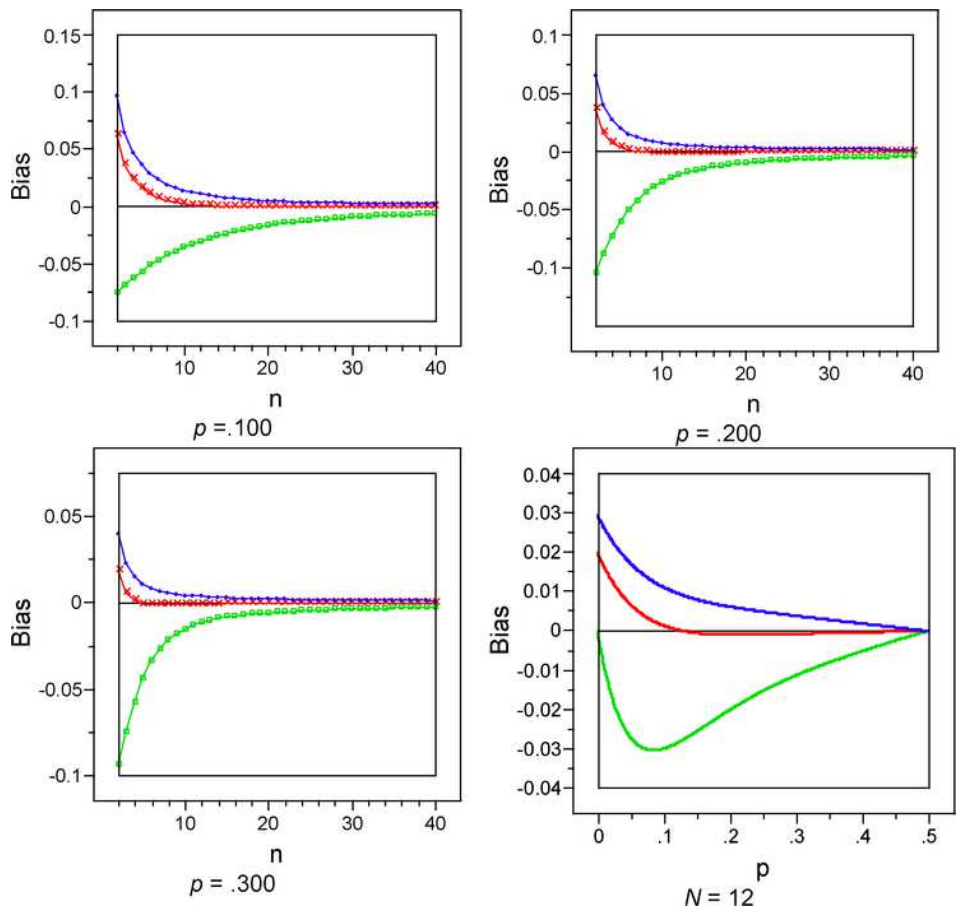

Fig. 1.  *Bias as defined in (2.7) for $Y^{(3/8)}$ (top curve), $Y^{(1/4)}$ (middle curve), $Y^{(0)}$ (bottom curve). Three plots show values of bias for various values of $N$ for $p = 0.100$, 0.200, 0.300, respectively. The 4th plot shows bias for $N = 12$ for various $p$.*

**3. Methodology, part II; estimation, prediction and validation.**  As discussed at (2.1), we will begin with a set of baseball batting records containing values generically denoted as $\{H_i, N_i\}$ for a sample of baseball players. These records may be for only part of a season and may consist of records for all the major league batters having a value of $H_i$ above a certain threshold, or may consist only of a sub-sample of such records.

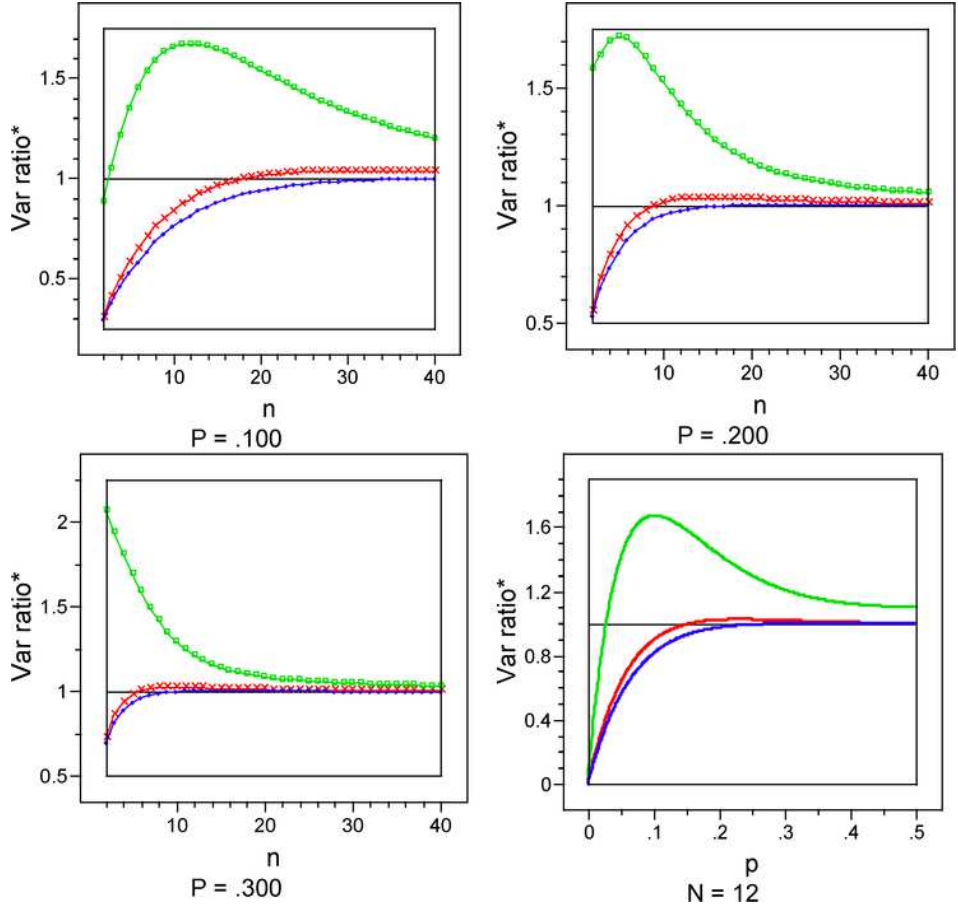

FIG. 2. *Variance ratio\* as defined in (2.8) for $Y^0$ (top curve), $Y^{(1/4)}$ (middle curve), $Y^{(3/8)}$ (bottom curve). Three plots show values of the ratio for various values of $N$ for $p = 0.100$, 0.200, 0.300, respectively. The 4th plot shows ratio for $N = 12$ for various $p$.*

In accordance with the discussion in Section 2, we will then write

$$(3.1) \qquad X_i = \arcsin\sqrt{\frac{H_i + 1/4}{N_i + 1/2}}, \qquad \theta_i = \arcsin\sqrt{p_i}.$$

We will assume that each $X_i$ is (approximately) normally distributed and they are all independent, a situation which we summarize by writing

$$(3.2) \qquad X_i \sim N(\theta_i, \sigma_i^2), \qquad \text{where } \sigma_i^2 = \frac{1}{4N_i}.$$

Much of the analysis that follows is grounded on the validity of this assumption; and to save space, we will proceed on that basis without further comment.

As the first concrete example, in Section 5 we will study records for each half season, denoted by $\{H_{ji}, N_{ji}\}$, $j = 1, 2$, $i = 1, \ldots, \mathcal{P}_j$. We assume

$$\theta_{ji} = \theta_i, \qquad j = 1, 2, \tag{3.3}$$

does not depend on the half of the season, $j$. In Section 7 we investigate empirical validity of such an assumption. Estimates for $\theta_i$ will be drawn from the values $\{X_{1i}, N_{1i} : i \in \mathcal{S}_1\}$ corresponding to the original observations $\{H_{1i}, N_{1i} : i \in \mathcal{S}_1\}$, where $\mathcal{S}_j = \{i : H_{ji} \geq 11\}$. As validation of the estimator, we compare the estimates to the corresponding observed value of $X_{2i}$. The validation is performed only over the set of indices $i \in \mathcal{S}_1 \cap \mathcal{S}_2$.

To fix the later terminology, let $\delta = \{\delta_i : i \in \mathcal{S}_1\}$ denote any estimator of $\{\theta_i : i \in \mathcal{S}_1\}$ based on $\{X_{1i}, N_{1i} : i \in \mathcal{S}_1\}$. Define the **S**um of **S**quared **P**rediction **E**rror as

$$SSPE[\delta] = \sum_{i \in \mathcal{S}_1 \cap \mathcal{S}_2} (X_{2i} - \delta_i)^2. \tag{3.4}$$

We will use the term "estimator" and "predictor" interchangeably for a procedure $\delta = \{\delta_i : i \in \mathcal{S}_1\}$, since it serves both purposes. It is desirable to adopt estimation methods for which SSPE is small.

The SSPE can serve directly as an estimate of the prediction error. It can also be easily manipulated to provide an estimate of the original estimation error. We will take the second perspective. Begin by writing the estimated squared error from the validation process as

$$SSPE[\delta] = \sum_{i \in \mathcal{S}_1 \cap \mathcal{S}_2} (\delta_i - X_{2i})^2 + \sum_{i \in \mathcal{S}_1 \cap \mathcal{S}_2} (X_{2i} - \theta_i)^2$$
$$- \sum_{i \in \mathcal{S}_1 \cap \mathcal{S}_2} 2(\delta_i - X_{2i})(X_{2i} - \theta_i).$$

The conditional expectation given $X_1$ of the third summand on the right is 0. For the middle term, observe that

$$E\left(\sum_{i \in \mathcal{S}_1 \cap \mathcal{S}_2} (X_{2i} - \theta_i)^2 \Big| X_1\right) = \sum_{i \in \mathcal{S}_1 \cap \mathcal{S}_2} \frac{1}{4N_{2i}}.$$

This yields as the natural estimate of the total squared error,

$$\widehat{TSE}[\delta] = SSPE[\delta] - \sum_{i \in \mathcal{S}_1 \cap \mathcal{S}_2} \frac{1}{4N_{2i}}.$$

In other words, $\widehat{TSE}[\delta] = SSPE[\delta] - E(SSPE[\theta])$, where $SSPE[\theta]$ denotes the sum of squared prediction error that would be achieved by an oracle who knew and used the true value of $\theta = \{\theta_i\}$.

For comparisons of various estimators in various situations, it is a little more convenient to re-normalize this. The naïve estimator $\delta_0(X) = X$ is a standard common-sense procedure whose performance will be investigated in all contexts. Because of this, a natural normalization is to divide by the estimated total squared error of the naïve estimator over the same set of batters. Thus, we define the normalized estimated squared error as

$$(3.5) \qquad \widehat{TSE}^*[\delta] = \frac{\widehat{TSE}[\delta]}{\widehat{TSE}[\delta_0]}.$$

In this way, $\widehat{TSE}^*[\delta_0] = 1$.

The estimators we adopt are primarily motivated by the normal setting in (3.2), so it seems statistically natural to validate them in that setting, as in (3.4)–(3.5). However, from the baseball context, it is more natural to consider prediction of the averages $\{R_{2i} : i \in \mathcal{S}_1 \cap \mathcal{S}_2\}$. For this purpose, given an estimation procedure $\tilde{R}$, the validation criteria become

$$\widehat{TSE}_R[\tilde{R}] = \sum_{i \in \mathcal{S}_1 \cap \mathcal{S}_2} (R_{2i} - \tilde{R}_i)^2 - \sum_{i \in \mathcal{S}_1 \cap \mathcal{S}_2} \frac{R_{2i}(1 - R_{2i})}{N_{2i}},$$

$$(3.6)$$

$$\widehat{TSE}_R^*[\tilde{R}] = \frac{\widehat{TSE}_R[\tilde{R}]}{\widehat{TSE}_R[\tilde{R}_0]},$$

where $\tilde{R}_0 = \{R_{1i}\}$.

In Sections 5 and 6 we compare the performance of several estimators as measured in terms of $\widehat{TSE}^*[\delta]$ and $\widehat{TSE}_R^*[\delta]$. See, for example, Table 2. An additional criterion, introduced in (5.1), is also investigated in that table. The following section contains definitions and motivations for the estimators whose performance will be examined.

## 4. Methodology, part III; description of estimators.

*Naïve estimator.* The simplest procedure is to use the first-half performance in order to predict the second half performance. Symbolically, this is the estimator

$$(4.1) \qquad \delta_0(X_{1i}) = X_{1i}.$$

*Overall mean.* Another extremely simple estimator is the overall mean. By using the overall mean this estimator ignores the first-half performance of each individual batter. Symbolically, this estimator is

$$(4.2) \qquad \delta(X_{1i}) = \bar{X}_1 = \mathcal{P}_1^{-1} \sum X_{1i}.$$

(For notational simplicity, we will usually use the symbol, $\delta$, for *all* our estimators, with no subscript or other identifier; but when necessary we will differentiate among them by name or by reference to their formula number.)

*Parametric empirical Bayes* (*method of moments*). The parametric empirical Bayes model for the current context originated with Stein (1962), followed by Lindley (1962). It is closely related to random effects models already familiar at the time, as discussed in Brown (2007), and can also be viewed as a specialization of the original nonparametric empirical Bayes formulation of Robbins (1951, 1956) that is described later in this section. The motivation for this estimator begins with supposition of a model in which

$$\text{(4.3)} \qquad\qquad \theta_i \sim N(\mu, \tau^2), \qquad \text{independent,}$$

where $\mu, \tau^2$ are unknown parameters to be estimated, and are often referred to as "hyper-parameters." If $\mu, \tau^2$ were known, then under (4.3) the Bayes estimator of $\theta_i$ would be

$$\text{(4.4)} \qquad\qquad \theta_i^{\text{Bayes}} = \mu + \frac{\tau^2}{\tau^2 + \sigma_{1i}^2}(X_{1i} - \mu),$$

where $\sigma_{ji}^2 = 1/4N_{ji}$,

Under the supposition (4.3) [and the normality assumption (3.2)], the observed variables $\{X_{1i}\}$ are marginally distributed according to

$$\text{(4.5)} \qquad\qquad\qquad X_{1i} \sim N(\mu, \tau^2 + \sigma_{1i}^2).$$

The empirical Bayes concept is to use the $\{X_{1i}\}$ distributed as in (4.5) in order to estimate $\mu, \tau^2$, and then to substitute the estimators of $\mu, \tau^2$ into Bayes formula, (4.4), in order to yield an estimate of $\{\theta_i\}$.

There are several plausible estimators for $\mu, \tau^2$ that can be used here. We present two of these as being of significant interest. As will be seen from Section 5, the performance of the resulting procedures differ somewhat, and this involves technical differences in the definitions of the procedures. The first estimator involves a Method of Moments idea based on (4.5). This requires iteratively solving a system of two equations, given as follows:

$$
\begin{aligned}
\text{(4.6)} \qquad & \tilde{\mu} = \frac{\sum X_{1i}/(\tilde{\tau}^2 + \sigma_{1i}^2)}{\sum 1/(\tilde{\tau}^2 + \sigma_{1i}^2)}, \\[2mm]
& \tilde{\tau}^2 = \frac{(\sum(X_{1i} - \tilde{\mu})^2 (-(\mathcal{P}_1 - 1)/\mathcal{P}_1) \sum \sigma_{1i}^2)_+}{\mathcal{P}_1 - 1}.
\end{aligned}
$$

As motivation for this estimator, note that if the positive-part sign is omitted in the definition, then one has the unbiasedness conditions $E(\tilde{\mu}) = \mu, E(\tilde{\tau}^2) = \tau^2$. The estimator of $\mu$ is chosen to be best-linear-unbiased. The positive-part sign in the definition of $\tilde{\tau}^2$ is a commonsense improvement on the estimator without that modification. Apart from this, the estimator of $\tilde{\tau}^2$ is not the only plausible unbiased estimate, and there are further motivations for the choice in (4.6) in terms of asymptotic Bayes and admissibility ideas, as discussed in Brown (2007).

[In practice, with data like that in our baseball examples, one iteration of this system yields almost the same accuracy as does convergence to a full solution. The one step iteration involves solving for the first iteration of $\tilde{\tau}^2$ by simply plugging $\bar{X}_1$. into (4.6) in place of $\tilde{\mu}$. Then this initial iteration for $\tilde{\tau}^2$ can be substituted into the first equation of (4.6) to yield a first value for $\tilde{\mu}$. When $\bar{X}_1$. is numerically close to $(\sum X_{1i}/\sigma_{1i}^2)/(\sum 1/\sigma_{1i}^2)$ this one-step procedure yields a satisfactory answer; otherwise, additional iterations may be needed to find a better approximation to the solution of (4.6).]

Symbolically, the parametric empirical Bayes (Method of Moments) estimator [**EB(MM)** as an abbreviation] can be written as

$$(4.7) \qquad \delta_i = \tilde{\mu} + \frac{\tilde{\tau}^2}{\tilde{\tau}^2 + \sigma_{1i}^2}(X_{1i} - \tilde{\mu}),$$

with $\tilde{\mu}, \tilde{\tau}^2$ as in (4.6).

*Parametric empirical Bayes* (*maximum likelihood*). Efron and Morris (1975) suggest the above idea, but with a modified maximum likelihood estimator in place of (4.6). We will investigate their maximum likelihood proposal, but, for simplicity, will implement it without the minor modification that they suggest. In place of (4.6) use the maximum likelihood estimators, $\hat{\mu}, \hat{\tau}^2$ based on the distribution (4.5). These are the solution to the system

$$(4.8) \qquad \hat{\mu} = \frac{\sum X_{1i}/(\hat{\tau}^2 + \sigma_{1i}^2)}{\sum 1/(\hat{\tau}^2 + \sigma_{1i}^2)},$$

$$\sum \frac{(X_{1i} - \hat{\mu})^2}{(\hat{\tau}^2 + \sigma_{1i}^2)^2} = \sum \frac{1}{\hat{\tau}^2 + \sigma_{1i}^2}.$$

Substitution in (4.4) then yields the parametric empirical Bayes (Maximum Likelihood) estimator [**EB(ML)** as an abbreviation]:

$$(4.9) \qquad \delta_i = \hat{\mu} + \frac{\hat{\tau}^2}{\hat{\tau}^2 + \sigma_{1i}^2}(X_{1i} - \hat{\mu}).$$

*Nonparametric empirical Bayes.* Begin with the weaker supposition than (4.3) that

$$(4.10) \qquad \theta_i \sim G, \qquad \text{independent,}$$

where $G$ denotes an unknown distribution function. If $G$ were known, then the Bayes estimator would be given by

$$(4.11) \qquad (\theta_G)_i = E(\theta_i|X) = \frac{\int \theta_i \varphi((X_{1i} - \theta)/\sigma_{1i}) G(d\theta)}{\int \varphi((X_{1i} - \theta)/\sigma_{1i}) G(d\theta)},$$

where $\varphi$ denotes the standard normal density.

The original empirical Bayes idea, as formulated in Robbins ([1951], [1956]), is to use the observations to produce an approximation to $\theta_G$, even though $G$ is not known. As Robbins observed, and others have noted in various contexts, it is often more practical and effective to estimate $\theta_G$ indirectly, rather than to try to use the observations directly in order to estimate $G$ and then substitute that estimate into (4.11). [But, we note that C. Zhang (personal communication) has recently described for homoscedastic data a feasible calculation of a deconvolution estimator for $G$ that could be directly substituted in (4.11).]

Brown and Greenshtein ([2007]) propose an indirectly motivated estimator. They begin with the formula from Brown ([1971]) which states that

$$(\theta_G(X_1))_i = X_1 + \sigma_{1i}^2 \frac{\frac{\partial g_i^*}{\partial X_{1i}}(X_1)}{g_i^*(X_1)},$$

(4.12)

$$\text{where } g_i^*(X_1) = \int \varphi((X_{1i} - \theta)/\sigma_{1i}) G(d\theta).$$

The next step is to estimate $g^*$ by a particular, generalized form of the kernel estimator preliminary to substitution in (4.12). The coordinate values of this kernel estimator depend on the values of $\sigma_{1i}^2$ and the kernel weights also depend on $\{\sigma_{1k}^2\}$.

Let $\sqrt{h} > 0$ denote the bandwidth constant for this kernel estimator. Then define

$$\tilde{g}_i^*(X) = \left( \sum_k \frac{I_{\{k:(1+h)\sigma_{1i}^2 - \sigma_{1k}^2 > 0\}}(k)}{\sqrt{(1+h)(\sigma_{1k}^2 \vee \sigma_{1i}^2) - \sigma_{1k}^2}} \right.$$

(4.13)

$$\left. \times \varphi((X_{1i} - X_{1k})/\sqrt{(1+h)(\sigma_{1k}^2 \vee \sigma_{1i}^2) - \sigma_{1k}^2}) \right)$$

$$\times \left( \sum_k I_{\{k:(1+h)\sigma_{1i}^2 - \sigma_{1k}^2 > 0\}}(k) \right)^{-1}.$$

Finally, define the corresponding nonparametric empirical Bayes estimator (**NPEB** as an abbreviation) as

$$\delta_i(X_1) = X_{1i} + \sigma_{1i}^2 \frac{\frac{\partial \tilde{g}_i^*}{\partial X_{1i}}(X_1)}{\tilde{g}_i^*(X_1)}.$$

(4.14)

To further motivate this definition, note that calculation under the assumption (4.10) yields that, for any fixed value of $x \in \Re$,

$$E\left( \frac{\varphi((x - X_{1k})/\sqrt{(1+h)(\sigma_{1k}^2 \vee \sigma_{1i}^2) - \sigma_{1k}^2})}{\sqrt{(1+h)(\sigma_{1k}^2 \vee \sigma_{1i}^2) - \sigma_{1k}^2}} \right)$$

(4.15)

$$= \int \frac{\varphi((x-\theta)/\sqrt{(1+h)(\sigma_{1k}^2 \vee \sigma_{1i}^2)})}{\sqrt{(1+h)(\sigma_{1k}^2 \vee \sigma_{1i}^2)}} G(d\theta).$$

The integrand in the preceding expression is a normal density with mean $\theta$ and variance $(1+h)(\sigma_{1k}^2 \vee \sigma_{1i}^2)$. The summation in (4.13) extend only over values of $\sigma_{1k}^2 \leq (1+h)\sigma_{1i}^2$. Then, since $h$ is small, for these values we have the approximation $(1+h)(\sigma_{1k}^2 \vee \sigma_{1i}^2) \approx \sigma_{1i}^2$, so that comparing (4.12) and (4.15) yields $\tilde{g}_i^* \approx g_i^*$. A similar heuristic approximation is valid for the partial derivatives that appear in (4.12) and (4.14). Hence, (4.14) appears as a potentially useful estimate of the Bayes solution (4.12).

In the applications below we used the value $h = 0.25$ for the situations where $\mathcal{P} > 200$ and $h = 0.30$ for the smaller subgroup having $\mathcal{P} = 81$. After taking into account the reduction in effective sample size in (4.13) as a result of the heteroscedasticity, this choice is consistent with suggestions in Brown and Greenshtein (2007) to use $h \approx 1/\log \mathcal{P}$. Performance of the estimators as described in Section 5 seemed to be moderately robust with respect to choices of $h$ within a range of about $\pm 0.05$ of these values.

*Harmonic Bayes estimator.* An alternate path beginning with the hierarchical structure (4.3) involves placing a prior distribution or measure on the hyper-parameter. One prior that has appealing properties in this setting is to let $\mu$ be uniform on $(-\infty, \infty)$ and to let $\tau^2$ be (independently) uniformly distributed on $(0, \infty)$. The resulting marginal distribution on $\theta \in \Re^{\mathcal{P}}$ involves the so-called harmonic prior. Specifically, $\psi = \theta - \bar{\theta}\mathbf{1} \in \Re^{\mathcal{P}-1}$ has density

$$f(\psi) \propto 1/\|\psi\|^{\mathcal{P}-3},$$

as can essentially be seen in Strawderman (1971, 1973). This prior density is discussed in Stein (1973, 1981), where it is shown that the resulting formal Bayes estimator for $\psi$ is minimax and admissible in the homoscedastic case. (In our context, this case is when all $N_i$ are equal.) However, even in the homoscedastic case, it is not true that the estimate of $\theta$ defined by the above prior is admissible. However, it is not far from being admissible, and the possible numerical improvement is very small. See Brown and Zhao (2007).

The expression for the posterior can be manipulated via operations such as change of variables and explicit integration of some interior integrals to reach a computationally convenient form for the posterior density of $\mu, \tau^2$. For notational convenience, let $\gamma = \tau^2$. Then the posterior density of $\mu, \gamma$ has the expression

(4.16) $\quad f(\mu, \gamma | X_1) \propto \left[ \prod (\gamma + \sigma_{1j}^2) \right]^{-1/2} \exp\left( -\sum \frac{(X_{1j} - \mu)^2}{2(\gamma + \sigma_{1j}^2)} \right).$

The (formal) harmonic Bayes estimator (**HB** as an abbreviation) is thus given by

$$\delta_i(X_1) = E\left(\mu + \frac{\gamma}{\gamma + \sigma_{1i}^2}(X_{1i} - \mu)\Big| X_1\right).$$

Evaluation of this estimator requires numerical integration of $\mathcal{P}_1 + 1$ double integrals.

[In practice, this computation was slightly facilitated by making the change of variables $\omega = \underline{\sigma}^2/(\tau + \underline{\sigma}^2)$, where $\underline{\sigma}^2 = \min\{\sigma_{1i}^2\} > 0$, and also by noting that the posterior for $\mu$ is quite tightly concentrated around the value at which its marginal density is a maximum.]

*James–Stein estimator.* For the present heteroscedastic setting in which shrinkage to a common mean is desired, the natural extension of the original James and Stein (1961) positive-part estimator has the form (**J–S**)

$$(4.17) \qquad \delta(X_1) = \hat{\mu}_1 + \left(1 - \frac{\mathcal{P}_1 - 3}{\sum(X_{1i} - \hat{\mu}_1)^2/\sigma_{1i}^2}\right)_+ (X_{1i} - \hat{\mu}_1),$$

where

$$\hat{\mu}_1 = \frac{\sum X_{1i}/\sigma_{1i}^2}{\sum 1/\sigma_{1i}^2}.$$

Note that this estimator shrinks all coordinates of $X_1$ by a common multiple (toward $\hat{\mu}_1$), in contrast to the preceding Bayes and empirical Bayes estimators. Brown and Zhao (2006) suggests modifying this estimator slightly, either by increasing the constant to $\mathcal{P}_1 - 2$ or by adding an extra (small) shrinkage term; but the numerical difference in the current context is nearly negligible, so we will use the traditional form, above.

REMARK (Minimaxity). The original positive-part estimator proposed in James and Stein (1961) is

$$\delta_{\mathrm{orig}}(X_1) = \left(1 - \frac{\mathcal{P}_1 - 2}{\sum(X_{1i} - \hat{\mu}_1)^2/\sigma_{1i}^2}\right)_+ X_1.$$

The estimator (4.17) is a natural modification of this that provides shrinkage toward the vector whose coordinates are all equal, rather than toward the origin. Such a modification was suggested in Lindley (1962) and amplified in Stein (1962), page 295, for the homoscedastic case, which corresponds here to the case in which all values of $N_{1i}$ are equal. The formula (4.17) involves the natural extension of that reasoning.

Stein's estimator was proved in James and Stein (1961) to be minimax in the homoscedastic case. [A more modern proof of this can be found in Stein

(1962, 1973) and in many recent textbooks, such as Lehmann and Casella
(1998).] It was also proved minimax for the heteroscedastic case under a
modified loss function that is directly related to the weighted prediction
criterion defined in (5.1) below. It is not necessarily minimax with respect
to un-weighted quadratic loss or to a prediction criterion such as (3.4) or
(3.5). However, in the situations at hand, it can be shown that the arrays of
values of $\{N_{1i}\}$ are such that minimaxity does hold. To establish this, reason
from Brown (1975), Theorem 3, or from more contemporary statements in
Berger (1985), Theorem 5.20, or Lehmann and Casella (1998), Theorem 5.7.

However, even though it is minimax, the J–S estimator need not provide
the most desirable predictor in situations like the present one. This is es-
pecially so if the values of $N_{1i}$ are not stochastically related to the batting
averages, $H_{1i}/N_{1i}$ (or if some relation exists, but it is not a strong one). It is
suggested in Brown (2007) that in such a case it may be more desirable to use
a procedure based on a spherically symmetric prior, such as the harmonic
Bayes estimator described above, or to use an empirical Bayes procedure
based on a symmetric assumption, such as (4.3). The current study does
not attempt to settle the theoretical issue of which forms of estimator are
generally more desirable in settings like the present one. But, we shall see
that for the data under consideration some of these estimators do indeed
perform better than the J–S estimator.

## 5. Prediction based on the first half season.

5.1. *All players, via* $\widehat{TSE}^*$. As described at (3.3), we divide the season
into two parts, consisting of the first three months and the remainder of
the season. We consider only batters having $N_{1i} \geq 11$, and use the results
for these batters in order to predict the batting performance of all of these
batters that also have $N_{2i} \geq 11$. The first data column of Table 2 gives the
values of $\widehat{TSE}^*$, as defined in (3.5), for the various predictors discussed in
the previous section. The remaining columns of the table will be discussed
in Section 5.2.

REMARKS. Here are some remarks concerning the entries for $\widehat{TSE}^*$:

1. The *worst* performing predictor in this column is the naïve predictor.
This predictor directly uses each $X_{1i}$ to predict the corresponding $X_{2i}$.

On the other extreme, prediction to the overall mean ignores the individ-
ual first-half performance of the batters (other than to compute the overall
mean). Even so, it performs *better* than the naïve predictor! (Overall means
do not change much from season to season. It would also considerably out-
perform the naïve estimator if one were to ignore first half behavior entirely,
and just predict all batters to perform according to the average of the first
half of the preceding season.)

TABLE 2
*Values for half-season predictions for all batters of $\widehat{TSE}^{*}$, $\widehat{TSE}_{R}^{*}$ and $\widehat{TWSE}^{*}$ [as defined in (5.1), below, and the discussion afterward]*

|  | All batters; $\widehat{TSE}^{*}$ | All batters; $\widehat{TSE}_{R}^{*}$ | All batters; $\widehat{TWSE}^{*}$ |
|---|---|---|---|
| $\mathcal{P}$ for estimation | 567 | 567 | 567 |
| $\mathcal{P}$ for validation | 499 | 499 | 499 |
| Naive | 1 | 1 | 1 |
| Group's mean | 0.852 | 0.887 | 1.120 (0.741[1]) |
| EB(MM) | 0.593 | 0.606 | 0.626 |
| EB(ML) | 0.902 | 0.925 | 0.607 |
| NP EB | 0.508 | 0.509 | 0.560 |
| Harmonic prior | 0.884 | 0.905 | 0.600 |
| James–Stein | 0.525 | 0.540 | 0.502 |

2. The best performing predictors in order are those corresponding to the nonparametric empirical Bayes method, the James–Stein method, and the parametric EB(MM) method. The performance of the parametric EB(ML) method and the true (formal) Bayes harmonic prior method is mediocre. They perform about equally poorly; indeed, the two estimators are numerically very similar, which is not surprising if one looks closely at the motivation for each.

3a. There are two explanations for the relatively poor performance of the EB(ML) and the HB estimators. First, Figure 3 contains the histogram for the values of $\{X_{1i}\}$. Note that this histogram is not well matched to a normal distribution. In fact, as suggested by the results in Table 1, it appears to be better modeled as a mixture of two distinct normal distributions. But the motivation for these two estimators involves the presumption in (4.3) that the true distribution of $\{\theta_i\}$ is normal, and this would entail that the $\{X_{1i}\}$ are also normally distributed. Hence, the situation in practice does not completely match well to the motivation supporting these estimators. However, this nonmatch is also true concerning the motivation for the EB(MM) and J–S estimators. Only the nonparametric EB estimator is designed to work well in situations where the $\{\theta_i\}$ are noticeably nonnormal. This provides the justification for the fact that the NPEB estimator performs best. The difference in performance between the EB(ML) or the HB estimator and the EB(MM) or J–S estimators apparently rests on a second respect in which the actual data is not well matched to the motivation for the estimators.

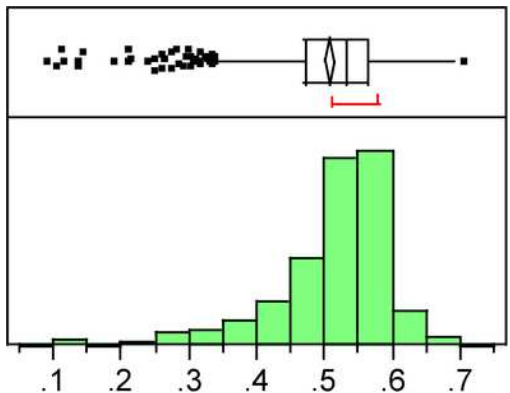

FIG. 3. *Histogram and box-plot for $\{X_{1i} : N_{1i} \geq 11\}$.*

3b. The second source of deviation from assumptions is that the sample values of $\{N_{1i}\}$ and $\{X_{1i}\}$ are moderately correlated. There is considerable correlation due to the fact that the pitchers generally have many fewer at-bats and much lower batting averages. (Their mean values for $N_1$ and $X_1$ are $\bar{X}_1 = 0.396, \bar{N}_1 = 25.1$, whereas for the nonpitchers the values are $\bar{X}_1 = 0.528, \bar{N}_1 = 157.8$.) Furthermore, even among the group consisting only of nonpitchers, there is also correlation. This correlation is evident from the following plot of $N_{1i}$ versus $X_{1i}$ for nonpitchers in $\mathcal{S}_1$. [Frey (2007) observed a qualitatively similar plot for the entire 2004 season.]

Although correlations as described above violate the basic assumptions motivating all of the empirical Bayes and the Bayes estimator, they seem to have a greater effect on the EB(ML) and the HB estimator than on the other three estimators under discussion. This effect manifests itself in terms of both the estimated mean and the estimate of $\tau^2$ which controls the shrinkage factor appearing in (4.4).

In the present situation, the more important effect is that on the estimate of $\tau^2$. Higher performing batters tend to have much higher numbers of at-bats. The EB(ML) estimator essentially computes a weighted estimate of $\tau^2$ with weights proportional to the values of $N_i$. It thus gives most weight to those higher average batters, whose averages are clustered closer together. The other EB estimators, such as EB(MM), essentially use an unweighted estimate of $\tau^2$, which results in a larger value for the estimate. A *smaller* estimate for $\tau^2$ results in an estimator which *shrinks more*. This results in performance that more closely resembles that of the overall mean, which is inferior to the more suitably calibrated shrinkage estimators, such as EB(MM) or NPEB.

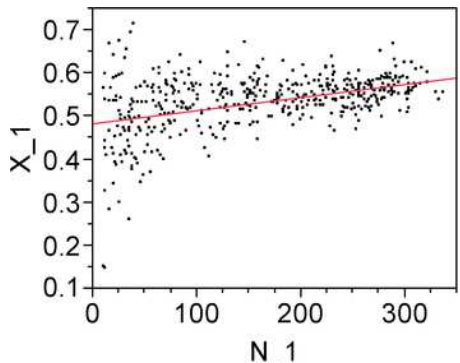

FIG. 4.   *Scatterplot of $X_1$ vs $N_1$ for nonpitchers. For this plot, $R^2 = 0.18$. (Overall, for all batters in $\mathcal{S}_1$, the value is $R^2 = 0.247$.)*

Figure 5 shows several of the estimators. Note that the EB(ML) estimator shrinks almost completely to $\hat{\mu}$. (The HB estimator, not shown, is very similar.) The J–S estimator is a linear function, and has a relatively steep slope. The NPEB estimator involves "shrinkage" in varying amounts (depending on the respective $N_i$) and toward somewhat different values depending on $X_i$. (Thus, some may feel that "shrinkage" is not a strictly correct term to describe its behavior.) Overall, there is considerable similarity between the NPEB and J–S estimators, and this is consistent with the fact that their overall performance is similar.

3c. In summary, the correlation between $\{N_{1i}\}$ and $\{X_{1i}\}$ is an important feature of the data. Such a correlation violates the statistical model that justifies our empirical Bayes and Bayes estimators. The results in Table 2 show for our data that, relative to the other estimators, the EB(MM) estimator and the HB estimator are not robust with respect to this type of deviation from the ideal assumptions. Further calculations (not reported here) also lead to the same conclusion in other, more general settings, that these estimators are not robust with respect to this type of deviation from assumptions, and should thus not be used if such deviations are suspected.

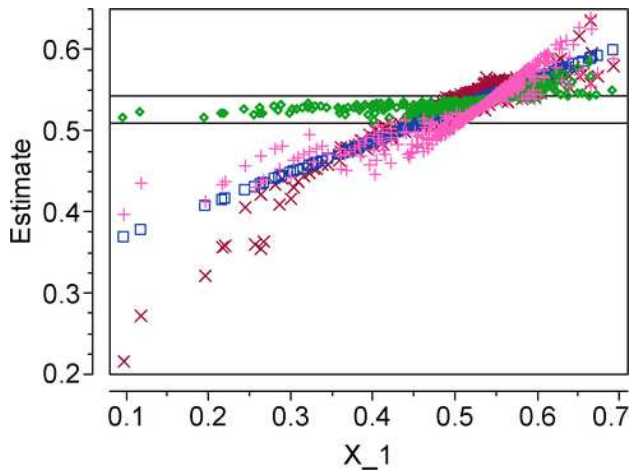

FIG. 5.  *Values of estimates as a function of $X_1$ for the full data set.* $\times = NPEB$, $\square = J$–$S$, $\diamond = EB(ML)$, $+ = EB(MM)$. *The lower horizontal line is $\bar{X}_1 = 0.509$, the upper one is $\hat{\mu} = 0.542$ of (4.8).*

### 5.2. *All players, other criteria.*

*via* $\widehat{TSE}_R^*$. $\widehat{TSE}_R^*$ as defined in (3.6) involves estimation of means for batting averages, rather than for values of $X$. The second data column of Table 3 contains the values of $\widehat{TSE}_R^*$ for all players.

For the naïve prediction here, we used just the first half batting average. The group mean used for the prediction here was the group mean of the first half averages. The other predictions used in the calculations for this column were derived by inverting the expression in the second part of (3.1); that is,

$$\tilde{R}_i = \sin^2 \hat{\theta}_i,$$

where the values of $\hat{\theta}_i$ were those used to derive $\widehat{TSE}^*$ in the first column of the table. Note that the results are very similar to those for $\widehat{TSE}^*$. This is a demonstration of the fact that prediction and validation can equally be carried out in terms of the $X$-values or in terms of actual batting averages. Because of this similarity, in the remainder of the paper we give only results for $X$-values since these are directly related to the motivation of the various estimators discussed in Section 4.

*All players, via a weighted squared-error criterion.* The prediction criteria described in Section 3 and studied above involve equal weights for all players. This type of criteria is suitable for some practical purposes, and is also a special focus for our study of the general performance of various estimators. For other practical purposes, it may be desirable to weight the performance of the predictors according to the number of at-bats of each player. This reflects a desire to concentrate on accuracy in predicting the performance of those batters who have the most at-bats. The most appropriate practical form of this criterion might be the one that weights squared prediction error according to the number of each player's second half at-bats. However, this number is unknown at the time of prediction, and so this criterion would involve an additional random quantity. For this reason, we prefer to study a prediction-loss that uses weights derived from the player's number of first half at-bats. Accordingly, the criterion used in Table 2 is

$$\widehat{TWSE}[\delta] = \sum_{i \in \mathcal{S}_1 \cap \mathcal{S}_2} N_{1i}(X_{2i} - \delta_i)^2 - \sum_{i \in \mathcal{S}_1 \cap \mathcal{S}_2} \frac{N_{1i}}{4N_{2i}},$$

(5.1)

$$\widehat{TWSE}^*[\delta] = \frac{\widehat{TWSE}[\delta]}{\widehat{TWSE}[\delta_0]}.$$

There are two table entries corresponding to the mean in this column. The first entry corresponds to the use of the ordinary sample mean as the predictor. The second entry, marked with the superscript [1] corresponds to the use of the weighted mean. The weighted mean is the generalized least square estimator relative to weighted squared error, so it is natural that its value of $\widehat{TWSE}^{*}$ should be considerably smaller than when using the unweighted mean. Its value is also considerably *less* than that for the naïve estimator.

The relative performance of the several estimators is somewhat different under this criterion than under $\widehat{TSE}^{*}$ and $\widehat{TSE}_{R}^{*}$. It is still the case that the naïve estimator performs *poorly*, and most of the other estimators are *better*. However, it is now the case that the J–S estimator performs the best. All the other estimators have rather similar performance under this criterion, with NPEB being slightly better than the others.

Remark 3, above, suggests that the correlation of $\{N_{1i}\}$ and $\{X_{1i}\}$ is related to the previously observed weaker performance of EB(ML) and the harmonic Bayes estimator. The use of $\widehat{TWSE}^{*}$ mitigates the effect of this correlation, since in the situation at hand it stresses accuracy of the prediction errors for the higher performing batters because these batters generally also have larger values of $N_{1i}$. It is also worth noting that the motivation for the James–Stein estimator involves exactly the sort of weighted squared error that appears in (5.1). Hence, it was to be anticipated that the J–S estimator would generally outperform the other estimators with respect to this criterion. [It is much more surprising to us that it also dominates EB(ML) and HB with respect to $\widehat{TSE}^{*}$ and $\widehat{TWSE}_{R}^{*}$.]

5.3. *Results for the two subgroups (nonpitchers and pitchers).* Remark 3b stresses that some of the performance characteristics of the estimators may be due to the correlation between the $\{N_{1i}\}$ and $\{X_{1i}\}$. This correlation is weaker or absent within the two subgroups of nonpitchers and of pitchers. If one looks only at the nonpitchers, then this correlation is somewhat weaker ($R^2 = 0.19$ vs $R^2 = 0.25$), and the sample distribution of $\{X_{1i}\}$ is somewhat closer to being normal. For the pitchers, the correlation is virtually zero ($R^2 = 0.0001$), and the sample distribution of $\{X_{1i}\}$ is close to normal. We might therefore expect some of the procedures —especially EB(ML) and the harmonic prior to have improved relative performance when used within these subgroups.

Table 3 contains values of the prediction criteria for predictors constructed separately from the first-half records of the nonpitchers and of the pitchers. There is considerable regularity in the relative performance of the estimators as compared with each other and only a few differences as compared with the pattern of results in Table 2.

For both subgroups it is still true that the naïve estimator has the *worst* performance, and here the overall mean is *much better*. Indeed, here it is not only much better, but with respect to $\widehat{TSE^*}$ none of the other estimators have significantly better performance, although only J–S is noticeably worse.

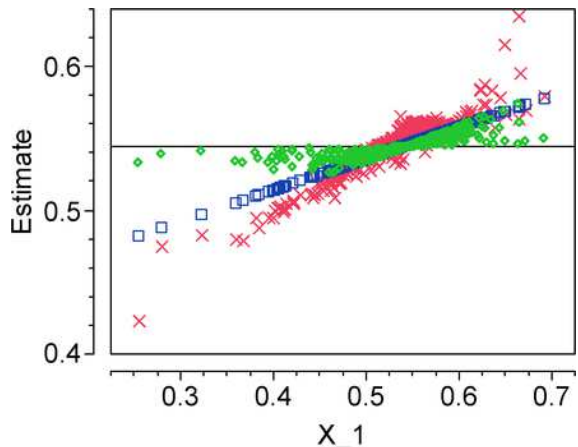

FIG. 6. *Values of estimates as a function of $X_i$ for the nonpitchers data set.* $\times = NPEB$, $\square = J$–$S$, $\diamond = EB(ML)$. *The horizontal line shows both* $\bar{X}_1 = 0.544$ *and* $\hat{\mu} = 0.546$ *of* (4.8).

For the subsample of nonpitchers, Figure 6 shows the estimators resulting from three of the procedures. We do not show the EB(MM) or HB estimators since these are quite similar to EB(ML) here, and all involve shrinkage almost to the sample mean. In fact, among the alternative estimators proposed in Section 4, all are comparable to the sample mean except for the James–Stein estimator, which has much worse performance. Figure 5 shows that the J–S estimator has very much less shrinkage than the other estimators, and is much more similar to the naïve estimator taking the values $\{X_{1i}\}$.

The relatively poor performance of the nonparametric EB estimator *wrt* $\widehat{TSE^*}$ for the subgroup of pitchers is perhaps related to the relatively small sample size. That estimator is constructed to perform well for moderate to large sample sizes, and perhaps the sample size here ($\mathcal{P}_1 = 81$) is somewhat marginal to get good performance for this estimator because of the presence of noticeable heteroscedasticity. (The sample values of $N_{1i}$ have a four-fold range, from 11 to 44.) Furthermore, the other (empirical) Bayes estimators are particularly constructed to perform well for situations where the values of $\{\theta_i\}$ are normally distributed, which appears to be very nearly the case here.

We found it somewhat surprising that the J–S estimator did not perform comparatively better for the subgroup of pitchers. Especially in the case of $\widehat{TWSE}^*$, all the motivating assumptions for the J–S estimator appear to hold quite closely, so one could expect it to perform very well. However, the estimators that outperform J–S (the parametric EB estimators and the HB estimator) are especially constructed to work well in the situation where the true values of $\{\theta_i\}$ are normally distributed, and that appears to be the case here. Even for weighted squared error ($\widehat{TWSE}^*$), these estimators retain their edge over J–S, which is especially designed for weighted squared error.

The weighting is not particularly relevant to the performance of these better performing estimators. This is because the weights (which derive from the $\{N_i\}$) are not particularly correlated with the values of the $\{\theta_i\}$.

*Simulations.* We performed some simulations to investigate the breadth of generality of the numerical results observed in Table 3. The focus of the present article is on the empirical results, rather than results of such simulations. Hence, we report only briefly on the nature of these simulation results insofar as they suggest the variability that one might expect from entries such as those in the tables.

We simulated results from the model (3.2)–(3.3) with arrays of values of $\{N_{1i}, N_{2i}\}$ taken from the actual data in the simulation, and used parameter values as suggested from the baseball data. In a second simulation we attempted to simulate from an ad-hoc model consistent with the type of correlation between the $\{N_{1i}\}$ and $\{X_{1i}\}$ as seen in Figure 4.

Overall, the actual results in Table 3 (as well as those in Table 2) are very consistent with the results from the simulations. In the simulations there is considerable variability of the magnitudes of the nonnormalized values of $\widehat{TSE}$. But there is much more stability in the normalized values, $\widehat{TSE}^*$, and in the relation between the entries in pairs of cells in the same column.

TABLE 3
*Values for half-season predictions for nonpitchers and for pitchers of $\widehat{TSE}^*$ and of weighted $\widehat{TWSE}^*$ [as defined in (5.1)]*

|  | Nonpitchers; $\widehat{TSE}^*$ | Nonpitchers; $\widehat{TWSE}^*$ | Pitchers; $\widehat{TSE}^*$ | Pitchers; $\widehat{TWSE}^*$ |
|---|---|---|---|---|
| $\mathcal{P}$ for estimation | 486 | 486 | 81 | 81 |
| $\mathcal{P}$ for validation | 435 | 435 | 64 | 64 |
| Naïve | 1 | 1 | 1 | 0.982 |
| Group's mean | 0.378 | 0.607 (0.561[1]) | 0.127 | 0.262 (0.262[1]) |
| EB(MM) | 0.387 | 0.494 | 0.129 | 0.191 |
| EB(ML) | 0.398 | 0.477 | 0.117 | 0.180 |
| NPEB | 0.372 | 0.527 | 0.212 | 0.266 |
| Harmonic prior | 0.391 | 0.473 | 0.128 | 0.190 |
| James–Stein | 0.359 | 0.469 | 0.164 | 0.226 |

(Superscript[1]: The numbers with superscript[1] are values relative to the weighted mean.)

For example, the pairwise differences from the simulation between the last six entries in the first column of the table had standard deviations in the simulation ranging from about 0.05 to 0.20. As a particular result, in the simulation from the model (3.2)–(3.3) (and with $\tau^2 = 0.0011$, which is consistent with the value seen in the baseball data) the difference between $\widehat{TSE}^*$ for the mean and for the J–S estimator had a mean value of 0.10 with a standard deviation of 0.09. This mean difference is of course considerably larger than that observed in the data, where the difference is only 0.019. But the observed difference is well within the range of values suggested by the simulation. Furthermore, the real-life situation has a correlation between the $\{N_{1i}\}$ and $\{X_{1i}\}$ which seems to affect the values of $\widehat{TSE}^*$ in Table 3, although only by additional amounts of a magnitude less than that of the already noted standard deviation. [As already remarked, the correlation has a greater effect on the behavior of EB(ML) and HB in Table 1.]

While not of great magnitude, these standard deviations are nevertheless large enough to cast doubt on whether the relations among the entries in the last six rows of the table would be stable across different baseball seasons.

The standard deviations for the analysis of pitchers were naturally noticeably larger. This is because the sample size there is only 81 as contrasted to 486 for the nonpitchers. It is also because the pitchers had more apparent variability in their values of $\{\theta_i\}$, and this was built into the parameter values used for the simulation.

The one conclusion that remains as being absolutely confirmed by the simulations is the inferiority of the naïve estimator relative to all the other estimators.

**6. Predictions based on other portions of the season.** The previous discussion involved producing estimates based on data from the first three months of the season. These estimates were then validated against the performance for the remainder of the season. It is possible to split the season in different fashions. For example, one can (try to) use data from the first month, and validate it against the performance for the remaining five months of the season. Or one could base predictions on the first five months on the season, and validate against performance in the last month. We discuss two such analyses below. In constructing such an analysis some care may need to be taken to guarantee similarity of the nature of batters in the estimation set and the validation set.

*Predictions based on one month of data.* In this situation the estimation set for all batters, $\mathcal{S}_1$, contains relatively few pitchers, but that is also true for the validation set $\mathcal{S}_1 \cap \mathcal{S}_2$. Hence, it is appropriate to conduct the validation study using all batters. Table 4 gives the results from this validation study.

The results reported in Table 4 are entirely consistent with earlier results and the previous discussion. By comparison with Table 2, the $\widehat{TSE}^*$ value here for the naïve estimator is very much larger than that for all the other estimators. This is because the naïve predictions based on only one-month's data are much less accurate than those based on three-month's data. On the other hand, the mean value for one month is not very different from that for three months. Hence, the mean has similar estimation accuracy in the setting of both Tables 2 and 4. In addition, as in Table 2, the value for EB(MM) is comparable to that for the overall mean, and the value for NPEB is noticeably better.

*Predictions based on five months of data.* If we use the first five months of the season as the portion on which to base predictions, then the validation set consists of the remainder of the season which is only slightly more than one month long. This results in an estimation sample that contains a hefty proportion of the low performing pitchers (102 pitchers and 532 nonpitchers). But the corresponding validation set contains relatively fewer pitchers (39 pitchers and 409 nonpitchers). For suitability of the type of validation study we are conducting, the validation set should resemble the estimation set in its important basic characteristics. Hence it is not useful to look at results here for all batters.

For this reason, we report only the results of an analysis based on non-pitchers with a five-month estimation set and a one-plus month validation set. Even here there is a problem concerning the structural similarity of the estimation and validation sets. Since the estimation set involves a much longer horizon, it contains a much larger proportion of rarely used (and low performing) batters. In order to attain better similarity in estimation and validation sets, we will require that the batters in the 5-month estimation set have values of $N_{1i} > 25$ to guarantee that they are not hitters who are extremely rarely used, and hence, unlikely to have at least 11 at-bats in the last month of the season.

With this type of estimation and validation situation, the values of $\{X_{1i}\}$ should be fairly good predictors of the corresponding $\{X_{2i}\}$, as suggested by considerations in the discussion following Table 6. But it is also true that the validation values of $\{X_{2i}\}$ are relatively close to each other so that the mean $\bar{X}_1$ is also a good predictor of the $\{X_{2i}\}$. The results in Table 5 are consistent with this. They are also consistent (if only marginally) with the discussion following (6.2), below, which suggests we should anticipate that $1 > \widehat{TSE}^*$[mean] for this type of study.

TABLE 4
*Values of $\widehat{TSE}^*$ for five estimators for prediction based on the first month for all batters*

| Naive | Mean | EB(MM) | NPEB | J–S |
|---|---|---|---|---|
| 1 | 0.250 | 0.240 | 0.169 | 0.218 |

*Naïve estimator vs group mean.* The fact that in our settings the overall group mean performs better than using the individual batting performances as predictors can be explained and amplified by a few simple calculations. The following table contains numerical sample quantities needed for these calculations.

In Table 6 the sums and sum of squared error (SSE) extend over $\mathcal{S}_1 \cap \mathcal{S}_2$ within the subgroup of the relevant row. Under the model (3.2)–(3.3),

$$\mathrm{Var}(X_{2i} - X_{1i}) = \frac{1}{4N_{2i}} + \frac{1}{4N_{1i}}.$$

Hence, the third data column of the table is the expectation of SSPE-naïve. Also,

$$(6.1) \quad E\left(\sum_{\mathcal{S}_1 \cap \mathcal{S}_2} (X_{2i} - \bar{X}_1)^2\right) = E(SSE_{\mathcal{S}_1 \cap \mathcal{S}_2}(X_2)) + E(\bar{X}_1 - E_{\mathcal{S}_1 \cap \mathcal{S}_2}(X_2)^2).$$

The second term on the right of (6.1) is numerically negligible. Hence, we have as a reasonably accurate approximation,

$$E\left(\sum_{\mathcal{S}_1 \cap \mathcal{S}_2} (X_{2i} - \bar{X}_1)^2\right) \approx SSE_{\mathcal{S}_1 \cap \mathcal{S}_2}(X_2).$$

TABLE 5
*Values of $\widehat{TSE}^*$ for five estimators based on the first 5 months, for nonpitchers (as described in the text)*

| Naive | Mean | EB(MM) | NPEB | J–S |
|---|---|---|---|---|
| 1 | 0.955 | 0.904 | 0.944 | 0.808 |

TABLE 6
*Statistics to evaluate ideal behavior of SSPE as defined in (3.4)*

| | $\sum_{\mathcal{S}_1 \cap \mathcal{S}_2} 1/4N_{1i}$ | $\sum_{\mathcal{S}_1 \cap \mathcal{S}_2} 1/4N_{2i}$ | Sum of prev. entries $= \mathbf{E(SSPE\text{-}naïve)}$ | $SSE_{\mathcal{S}_1 \cap \mathcal{S}_2}(X_2)$ $\approx \mathbf{E(SSPE\ to\ mean)}$ |
|---|---|---|---|---|
| All batters | 1.800 | 1.766 | 3.566 | 3.255 |
| Nonpitchers | 1.154 | 1.189 | 2.343 | 1.569 |
| Pitchers | 0.646 | 0.577 | 1.223 | 0.672 |

These are the entries in the last column of Table 6. These entries are all smaller than those in the preceding column. This shows that one should expect the data in Tables 2 and 3 to have SSPE[naïve] < SSPE[mean], which is equivalent to

(6.2)                                  $1 > $ SSPE[naïve].

We can also use this information to give some idea how much initial season data would be needed so that SSPE[naïve] $\approx$ SSPE[mean]. Multiplying the values of $N_{1i}$ by a constant factor, $c$, will multiply the values in the first data column of Table 4 by $1/c$. Hence, in order to have SSPE[naïve] $\approx$ SSPE[mean] over the full season, we need

$$c \approx \frac{1.800}{3.255 - 1.766} = 1.2.$$

In other words, an initial period of about $1.2 \times 3 = 3.6$ months should be enough to have SSPE[naïve] $\approx$ SSPE[mean] for the set of all players. (Actually, somewhat more than 3.6 months would probably be needed because adding additional time would bring some additional batters having small values of $N$ into the data under evaluation.)

For the subgroups, much more additional data would be needed, since these subgroups are much more homogeneous than the combined set of all players. The corresponding values of $c$ are as follows:

$$\text{for the nonpitchers } c \approx \frac{1.154}{1.569 - 1.189} = 3.0$$

and

$$\text{for the pitchers } c \approx \frac{0.646}{0.672 - 0.577} = 6.8.$$

Hence, one would need about $3/2$ seasons of initial data before a nonpitcher's initial batting average would overall be a better predictor of future performance than would the general mean value for all nonpitchers. For pitchers, one would need $3.4 = 6.8/2$ seasons for this same situation; so more than three years of data would be needed, and it would be necessary to assume that the pitcher's (latent) batting ability was stationary over this considerable time span.

**7. Validation of the independent binomial assumption.** The distributional assumption (2.1) states that each player's averages for subsequent seasonal periods can be modeled as independent binomial variables. Further, the mean-parameter, $p$, depends only on the player and does not change over successive periods of the season. Under discussion here are seasonal periods such as half-seasons, or somewhat shorter periods, such as successive one-month periods.

This assumption could be violated in several ways. The most prominent way would be if the player's latent batting ability ($p$) shifts systematically from period to period, as, for example, might happen with a batter whose abilities improve as the season progresses. For monthly periods it could also occur with a batter whose abilities are highest in the middle of the season and lower at the beginning and end.

A second mechanism that could lead to noticeable violation of these assumptions would be the existence of an intrinsically "streaky" batter. Such a batter is one whose true (but unknown) value of $p$ is higher for some substantial intervals of time within the basic period, and lower for a subsequent stretch of time. If these streaks are of a substantial length of time (say, one to two weeks) but still much less than that of the basic period under consideration (say, a month), then it could be that the player's mean latent ability during each period is (approximately) constant. However, such streakiness could result in violation of the binomial distribution assumption. The usual direction of such a violation would be in the statistically familiar direction of "over-dispersion." In this case the averages for each period could have constant mean values, but could have a variance that is larger than that given from the binomial distribution assumption. The issue of streakiness has been frequently discussed. See, for example, Albright (1993) in the baseball context or Gilovich, Vallone and Tversky (1985) for a discussion involving the sport of basketball.

Section 2 discusses the fact that under the assumption (2.1) the variables

$$X_{ji} = \sin\sqrt{\frac{H_{ji} + 1/4}{N_{ji} + 1/2}}$$

can be accurately treated as independent random variables having the distribution

$$X_{ji} \sim N(\sin\sqrt{p_i}, 1/4N_{ji}),$$

so long as $N_{ji} \geq 12$. This enables construction of a test for the null hypothesis (2.1) that has some sensitivity to detect nonconstant values of $p_i$ as a function of period, $j$, or deviations such as over-dispersion from the binomial distribution shape described in (2.1). Tests of this nature for a Poisson distribution have been discussed in Brown and Zhao (2002).

*Testing two halves of the season.* For the case where there are only two periods, $j = 1, 2$, one may look at the values of

$$(7.1) \qquad Z_i = \frac{X_{1i} - X_{2i}}{\sqrt{1/4N_{1i} + 1/4N_{2i}}}.$$

Under the null hypothesis, these values should be (very nearly) independent standard normal variables. One may use a normal quantile plot to graphically investigate whether this is the case, and any of several standard tests of normality (with mean 0 and variance 1) to provide $P$-values. Figure 7 shows the result of applying this test with the two periods being the first half and the second half of the season. The test was applied only for records for which $N_{ji} \geq 12, j = 1, 2$. There were 496 such records.

Figure 7 indicates that the values of $\{Z_i\}$ come close to attaining the desired standard normal distribution. There are no large outliers present, which suggests that there were no large scale shifts in individual ability between the two half seasons. Some deviation from normality is nevertheless evident, and this may be evidence of statistically significant, albeit small, deviation from the ideal binomial model. The $P$-value for the test of normality using the conventional Kolmogorov–Smirnov test is $P = 0.046$, only very slightly below the conventional value for identifying situations of possible interest.

*FDR procedure.* One may also apply an FDR procedure to this data. See Benjamini and Hochberg (1995) (B&H) and also Efron (2003). Let us follow B&H and let $q^*$ denote the False Discovery Rate, where a "discovery" corresponds to a statement that a certain batter's records appear not to be binomially distributed with constant $p_i$. Suppose $\{P_i : i = 1, \ldots, m\}$ is a collection of $P$-values corresponding to $m$ independent tests of hypotheses. Let $\{P_{(i)}\}$ denote the ordered values from smallest to largest, and let $\{H_{(i)}\}$ denote the corresponding null hypotheses. Let

$$k^* = k^*(q^*) = \max\left\{i : P_{(i)} \leq \frac{i}{m} q^*\right\}.$$

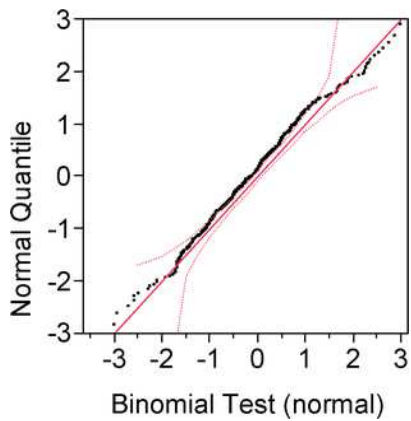

FIG. 7. *Normal quantile plot (mean 0 variance 1) for the values of $\{Z_i\}$ in (7.1).*

The procedure in B&H declares a "discovery" of the alternative hypothesis corresponding to $H_{(i)}$ for every $i \leq k^*$. Under this procedure, the expected proportion of false discoveries in the sense of B&H is at most $q^*$.

When the B&H procedure is used on the half-yearly data there will be no discoveries noted in this data when $q^*$ is set at the conventional level of $q^* = 0.05$. Indeed, there will be no discoveries noted until $q^*$ reaches nearly 0.5. At that value there will be 4 discoveries corresponding to the 4 largest values of $|Z_i|$—but, of course, by definition of the FDR procedure, one will then expect half of these discoveries to be false discoveries. As before, the overall conclusion here is that there is very little—if any—indication in the data that the assumption (2.1) is invalid for any individual batter.

*Testing with month-long periods.* It is also possible to use the same basic idea with more than two periods. A natural division for the data at hand is to construct a test based on 6 periods corresponding to the months of the season (with the last period including records from both September and October). To formally express the procedure, let the subscript $j = 1, \ldots, 6$ index the 6 months of data and let

$$m_i = \#\{j = 1, \ldots, 6 : N_{ji} \geq 12\}.$$

Then define

(7.2) $$Z_i^2 = \sum_{j \ni N_{ji} \geq 12} 4 N_{ji} (X_{ji} - \hat{X}_{\cdot i})^2 \qquad \text{where } \hat{X}_{\cdot i} = \frac{\sum_{j \ni N_{ji} \geq 12} N_{ji} X_{ji}}{\sum_{j \ni N_{ji} \geq 12} N_{ji}}.$$

Under the assumptions (3.2)–(3.3) which follows from (2.1), it will be the case that the random variables $Z_i^2$ are independent $\chi_{m_i-1}^2$ variables. Only indices $i$ for which $m_i \geq 2$ are of interest here, and so we will assume, *wlog*, that the batters of interest are indexed with indices $i = 1, \ldots, \mathcal{P}$. Here the effective sample size is $\mathcal{P} = 514$. (Only 36 of these 514 are pitchers, and their exclusion from the following analyses would have only very minor effects on the conclusions.)

There are several possible ways to collectively test the resulting null hypothesis that $Z_i^2 \sim \chi_{m_i-1}^2, i = 1, \ldots, \mathcal{P}$. The method we adopt here is to begin by defining

$$U_i = F_{m_i-1}^{\chi^2}(X_i),$$

where $F^{\chi^2}$ denotes the chi-squared CDF with the indicated degrees of freedom. Under the null hypothesis, the $\{U_i\}$ will be uniformly distributed.

In order to better display the data graphically, we will instead look at the values of $\Phi^{-1}(U_i)$. Under the null hypothesis, these values will be normally distributed. Figure 8 shows a normal quantile plot of the $\{\Phi^{-1}(U_i)\}$. Under the null hypothesis, this plot will demonstrate the ideal normal distribution pattern (nearly a 45° straight line). Under alternatives corresponding to streaks of lengths approximating a month or longer, the $\{U_i\}$ will be stochastically larger. In the case of streaks of shorter length, it could also be possible for the $\{U_i\}$ to be stochastically smaller. However, there is absolutely no evidence *in the monthly data* that such a phenomenon occurs with a strength or/and regularity to be visible in the analysis. Hence, we will concentrate in the following discussion on a one-sided test that rejects for large values of $\Phi^{-1}(U_i)$. (These, of course, correspond exactly to large values of $U_i$.) The test is thus attempting to detect streaks of the order of length nearly a month, or longer.

The (one-sided) $P$-value for this fit satisfies $P \geq 0.2$ for several plausible test statistics we calculated, such as the one-sided version of the Kolmogorov–Smirnov test. However, for a family-wise error-rate multiple comparison test of the null hypothesis that each $\Phi^{-1}(U_i)$ is standard normal (versus a one sided alternative), the family-wise $P$-value is $P^* = 0.055$, which is nearly significant at the conventional level of 0.05. More precisely,

$$P^* = 1 - \left[\max_i \Phi^{-1}(U_i)\right]^{514} = 1 - 0.99988922^{514} = 0.055.$$

Thus, this largest observation can be declared as a "discovery" at any FDR rate $q^* > 0.055$. There are no other FDR discoveries in the data until $q^*$ gets much larger. Even at $q^* = 0.5$, there are only two discoveries, corresponding to the two right-most points in Figure 8.

Out of curiosity, and to see some performance patterns that can qualify as a possibly streaky hitter (at the level of month-long streaks) in the presence of the random binomial noise implied by the model, we list the monthly records of these two players in Table 7, with the most significant batter listed first.

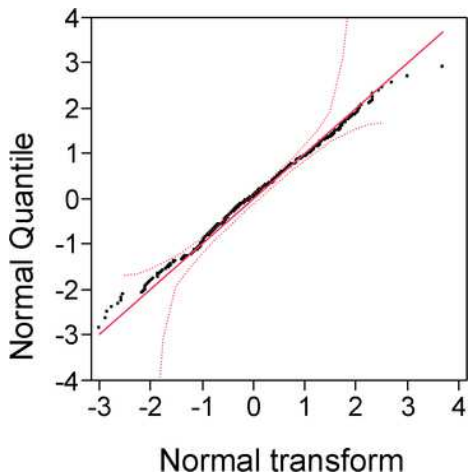

Fig. 8.    *Normal* $(0,1)$ *quantile plot for* $\{\Phi^{-1}(U_i)\}$.

TABLE 7
*Monthly records of the two hitters with largest value of $\Phi^{-1}(U)$*

| Batter | | Month 4 | M. 5 | M. 6 | M. 7 | M. 8 | M. 9–10 | Season |
|--------|-----|---------|------|------|------|------|---------|--------|
| Izturis | AB | 102 | 117 | 86 | 69 | 70 | – | 444 |
| | H | 34 | 41 | 9 | 17 | 13 | – | 114 |
| | pct | 0.333 | 0.350 | 0.105 | 0.246 | 0.186 | – | 0.257 |
| Crede | AB | 79 | 84 | 80 | 69 | 58 | 62 | 432 |
| | H | 24 | 13 | 22 | 21 | 6 | 23 | 109 |
| | pct | 0.304 | 0.155 | 0.275 | 0.304 | 0.103 | 0.371 | 0.252 |

## APPENDIX: SOME HITTERS DO EXHIBIT STREAKINESS OVER A SHORTER TIME SPAN

Sections 5 and 6 study the estimation of individual batting averages. As a partial validation of the estimation procedures under consideration, Section 7 constructs a test of streakiness in batting average at the level of half-year-long streaks or month-long streaks. Periods of a month or longer are the lengths of time relevant for the estimation procedures studied there. In brief, Section 7 of the paper finds no convincing evidence of any hitting streaks, or streakiness, at this level of granularity.

The same technique for investigating streakiness can be employed to examine whether there are streaks of shorter duration. The present postscript explores this issue, and finds convincing evidence of the existence of batting streaks lasting on the order of length of ten days (or longer).

**Construction of the test.** As before, the 2005 season is divided into segments. Here the segments will be 10 calendar days long except for the three days of the "All-Star Break," when no regular season games are played. The segment involving that break has ten days of scheduled regular games, running from July 2, 2005, through July 14, 2005.

As in the main article, let $N_{ji}, j = 1, \ldots, 18, i = 1, \ldots,$ denote the number of qualifying at-bats of player $i$ in period $j$. Let $H_{ji}$ denote the corresponding number of hits and

$$X_{ji} = \arcsin \sqrt{\frac{H_{ji} + 1/4}{N_{ji} + 1/2}},$$

as in (2.2). In order to eliminate pitchers and other batters who play only occasionally, we include in the analysis only players with a total of at least 100 at bats in the season. For reasons discussed in Sections 2 and 3 we wish to consider only the qualifying periods for each of these batters, where for batter $i$ the qualifying periods are defined as

$$\mathbb{Q}_i = \{j : N_{ji} \geq 12\}.$$

Then, as in Section 7, let

$$m_i = \#\{j = 1, \ldots, 18 : j \in \mathbb{Q}_i\}$$

and delete from the sample any batters with $m_i < 2$. (There were only two such batters among those who batted at least 100 times in the season.) The batters still in the sample can be labeled with subscripts $i = 1, \ldots, \mathcal{P} = 419$.

In order to describe the relevant notion of streakiness for the 10 day period here, consider the statistical model under which $H_{ji} \sim Bin(N_{ji}, p_i)$, independent. The *null* hypothesis that a batter's performance is *not* streaky is

$$\mathbf{H}_{0i} : p_{ji} = p_i \qquad \forall j \in \mathbb{Q}_i.$$

An identifiable violation of the null hypothesis indicates streaky performance by the batter.

As in (7.2), let

$$\text{(A.1)} \quad Z_i^2 = \sum_{j \ni N_{ji} \geq 12} 4 N_{ji} (X_{ji} - \hat{X}_{\cdot i})^2 \qquad \text{where } \hat{X}_{\cdot i} = \frac{\sum_{j \ni N_{ji} \geq 12} N_{ji} X_{ji}}{\sum_{j \ni N_{ji} \geq 12} N_{ji}}.$$

The tests suggested in Section 7 are based on

$$U_i = F^{\chi^2}_{m_i - 1}(Z_i^2),$$

or, equivalently, on $\Phi^{-1}(U_i)$. Large values of these statistics are significant. Accordingly, the $P$-value for batter $i$ is defined as

$$P_i = 1 - U_i.$$

The FDR procedure for discovering potentially streaky hitters (those who do not satisfy $\mathbf{H}_{0i}$) involves choosing a critical level $q^*$ and defining

$$k^* = k^*(q^*) = \max \left\{ i : P_{(i)} \leq \frac{i}{m} q^* \right\}.$$

Here, $\{P_{(i)}\}$ are the ordered $P$-values. The batters with $P_i \leq P_{(k^*)}$ are labeled as "discoveries" of those with potentially streaky performance. Benjamini and Hochberg (1995) show that the expected proportion of false discoveries is at most $q^*$.

This type of procedure was applied in Section 7 with the season divided into periods of two half-seasons, and also with 6 periods corresponding (nearly) to calendar months. In both those analyses no discoveries were declared at level $q^* = 0.05$. In the monthly analysis one discovery could be declared at a slightly larger level (the batter's name is C. Izturis; see Table 7 and also below), and no more until the level increased to above $q^* = 0.3$.

The results with 10-day periods are quite different. At $q^* = 0.05$, there are 32 discoveries among the 419 candidate players. Several of these 32 discoveries are familiar, regular players. Indeed, among the discoveries the modal value of $m_i$ is the maximum of 18 (9 players) and only 5 of the 32 have values $m_i \leq 9$.

Figure 9 is a normal quantile plot of $\Phi^{-1}(U_i)$ for all players. Under the null hypothesis that all $\mathbf{H}_{0i}$ are true, one would expect to see a (nearly) straight line.

It seems clear from this analysis that some players exhibited "streakiness" as measured by performance aggregated to 10 day time spans. (A more accurate, if less convenient substitute terminology for "streakiness" here would be "variability in latent performance ability.") There can be many potential explanations for such a finding. [Some possibilities could be runs of favored/unfavored pitchers and/or opposing teams, injury status for period(s) of the season, personal issues, just plain "streakiness," etc.] We will not attempt here to delve further to try to examine patterns of performance statistics that might lead to belief in one or the other of these explanations.

In order to display the type of performance(s) that can be classified as streaky we include some time series plots for a selection of batters classified as "discoveries" at $q^* = 0.05$. These plots are in terms of their ordinary

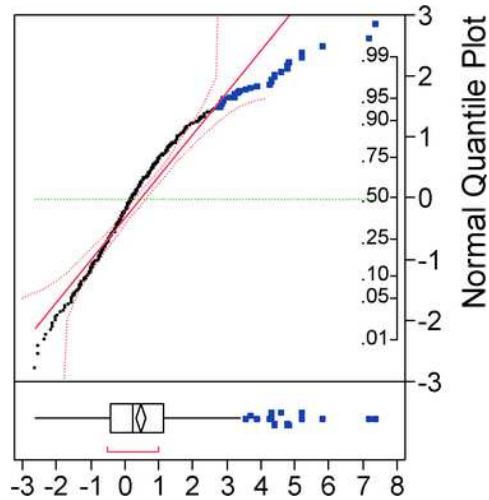

FIG. 9. Normal quantile plot of $\Phi^{-1}(U_i)$. The bold points correspond to values noted as discoveries at $q^* = 0.05$.

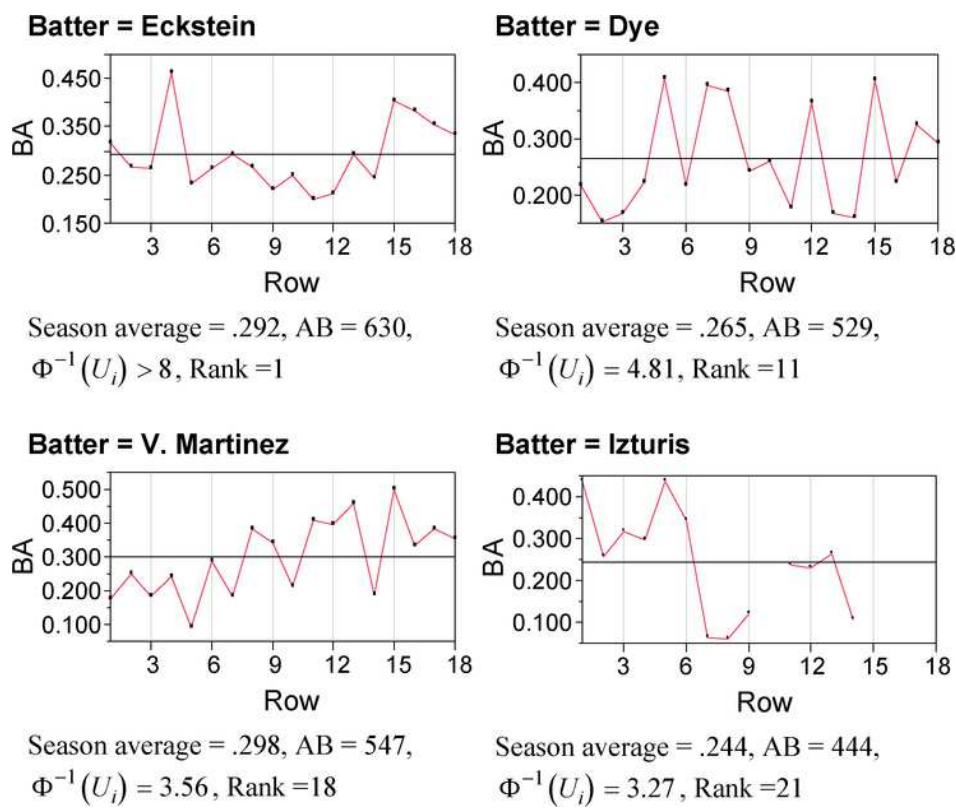

<image>Batter = Eckstein
Season average = .292, AB = 630,
$\Phi^{-1}(U_i) > 8$, Rank =1

Batter = Dye
Season average = .265, AB = 529,
$\Phi^{-1}(U_i) = 4.81$, Rank =11

Batter = V. Martinez
Season average = .298, AB = 547,
$\Phi^{-1}(U_i) = 3.56$, Rank =18

Batter = Izturis
Season average = .244, AB = 444,
$\Phi^{-1}(U_i) = 3.27$, Rank =21</image>

Fig. 10. *10 day averages of selected batters identified as "discoveries."*

batting average for each ten day period. We also include their season average (based only on qualifying periods), their total number of at bats for the season, their value of $\Phi^{-1}(U_i)$, and the rank of this value within all the 412 qualifying batters. In interpreting these plots, recall that the values of $U_i$ involve the players values of $N_{ji} = 1, \ldots, 18$ (which are not shown), along with the displayed values of their batting averages. Note also that the Y-axes are not all labeled consistently.

Albert ([2008](#)) uses a different methodology in an attempt to identify streakiness at a much finer level of duration than 10 days. He identifies two sets of top ten most streaky batters using two different statistical techniques. There is not much overlap between his top ten lists, nor between his lists and the 32 players we identified as "discoveries", as described above. Only C. Izturis and V. Martinez appear on both of his lists and among our 32 discoveries.

**Acknowledgments.** I would like to thank several colleagues for very enlightening discussion of earlier drafts of this paper, including the suggestion of several references cited in the references. Colleagues who were especially helpful include K. Shirley (who also prepared the original data set used here), S. Jensen, D. Small, A. Wyner and L. Zhao.

## SUPPLEMENTARY MATERIAL

**Major league batting records for 2005** (doi: 10.1214/07-AOAS138supp; .zip). The file gives monthly batting records (AB and H) for each Major League baseball players for the 2005 season. The names of the players are given, as well as a designation as to whether the player is a pitcher or not a pitcher.

## REFERENCES

ALBERT, J. (2008). Streaky hitting in baseball. *J. Quantitative Analysis in Sports* **4** Article 3. Available at http://www.bepress.com/jqas/vol4/iss1/3.

ALBERT, J. and BENNETT, J. (2001). *Curve Ball: Baseball, Statistics, and the Role of Chance in the Game.* Copernicus, New York. MR1835464

ALBRIGHT, S. C. (1993). A statistical analysis of hitting streaks in baseball (with discussion). *J. Amer. Statist. Assoc.* **88** 1175–1183.

ANSCOMBE, F. J. (1948). The transformation of Poisson, binomial and negative binomial data. *Biometrika* **35** 246–254. MR0028556

BARTLETT, M. S. (1936). The square root transformation in the analysis of variance. *J. Roy. Statist. Soc. Suppl.* **3** 68–78.

BARTLETT, M. S. (1947). The use of transformations. *Biometrics* **3** 39–52. MR0020763

BENJAMINI, Y. and HOCHBERG, Y. (1995). Controlling the false discovery rate: A practical and powerful approach to multiple testing. *J. R. Stat. Soc. Ser. B Stat. Methodol.* **57** 289–300. MR1325392

BERGER, J. O. (1985). *Statistical Decision Theory and Bayesian Analysis*, 2nd ed. Springer, New York. MR0804611

BROWN, L. D. (1971). Admissible estimators, recurrent diffusions and insoluble boundary value problems. *Ann. Math. Statist.* **42** 855–903. MR0286209

BROWN, L. D. (1975). Estimation with incompletely specified loss functions (the case of several location parameters). *J. Amer. Statist. Assoc.* **70** 417–427. MR0373082

BROWN, L. D. (2007). *Lecture Notes on Shrinkage Estimation.* Available at http://www-stat.wharton.upenn.edu/˜lbrown/.

BROWN, L. D. (2008). Supplement to "In-season prediction of batting averages: A field test of empirical Bayes and Bayes methodologies." DOI: 10.1214/07-AOAS138SUPP.

BROWN, L. D. and GREENSHTEIN, E. (2007). Empirical Bayes and compound decision approaches for estimation of a high dimensional vector of normal means. Manuscript.

BROWN, L. D. and ZHAO, L. (2002). A new test for the Poisson distribution. *Sankhyā Ser. A* **64** 611–625. MR1985402

BROWN, L. D. and ZHAO, L. (2006). Estimators for Gaussian models having a block-wise structure. Available at http://www-stat.wharton.upenn.edu/˜lbrown/.

BROWN, K. D. and ZHAO, L. (2007). A unified view of regression, shrinkage, empirical Bayes, hierarchical Bayes, and random effects. Available at http://www-stat.wharton.upenn.edu/˜lbrown/.

BROWN, L. D., CAI, T., ZHANG, R., ZHAO, L. and ZHOU, H. (2007). The root-unroot algorithm for density estimation as implemented via wavelet block thresholding. Available at http://www-stat.wharton.upenn.edu/˜lbrown/.

EFRON, B. (2003). Robbins, empirical Bayes and microarrays. *Ann. Statist.* **31** 366–378. MR1983533

EFRON, B. and MORRIS, C. (1975). Data analysis using Stein's estimator and its generalizations. *J. Amer. Stat. Assoc.* **70** 311–319. MR0391403

EFRON, B. and MORRIS, C. (1977). Stein's paradox in statistics. *Scientific American* **236** 119–127.

FREEMAN, M. F. and TUKEY, J. W. (1950). Transformations related to the angular and the square root. *Ann. Math. Statist.* **21** 607–611. MR0038028

FREY, J. (2007). Is an .833 hitter better than a .338 hitter? *The American Statistician* **61** 105–111. MR2368098

GILOVICH, T., VALLONE, R. and TVERSKY, A. (1985). The hot hand in basketball: On the misperception of random sequences. *Cognitive Psychology* **17** 295–314.

JAMES, W. and STEIN, C. (1961). Estimation with quadratic loss. *Proc. 4th Berkeley Symp. Probab. Statist.* **1** 367–379. Univ. California Press, Berkeley. MR0133191

LEHMANN, E. and CASELLA, G. (1998). *The Theory of Point Estimation*, 2nd ed. Springer, New York. MR1639875

LEWIS, M. (2004). *Moneyball*: *The Art of Winning an Unfair Game*. W. W. Norton, New York.

LINDLEY, D. (1962). Discussion on Proffesor Stein's paper. *J. Roy. Statist. Soc. B* **24** 285–287.

MOSTELLER, F. and YOUTZ, C. (1961). Tables of the Freeman–Tukey transformation for the binomial and Poisson distribution. *Biometrika* **48** 433–450. MR0132623

ROBBINS, H. (1951). An empirical Bayes approach to statistics. *Proc. 3rd Berkeley Symp. Math. Statist. Probab.* **1** 157–163. Univ. California Press, Berkeley. MR0084919

ROBBINS, H. (1956). Asymptotically subminimax solutions of compound statistical decision problems. *Proc. 2nd Berkeley Symp. Math. Statist. Probab.* 131–148. Univ. California Press, Berkeley. MR0044803

STEIN, C. (1962). Confidence sets for the mean of a multivariate normal distribution (with discussion). *J. Roy. Statist. Soc. Ser. B* **24** 265–296. MR0148184

STEIN, C. (1973). Estimation of the mean of a multivariate normal distribution. In *Proc. Prague Symp. on Asymptotic Statistics* **II** (*Charles Univ., Prague, 1973*) 345–381. Charles Univ., Prague. MR0381062

STEIN, C. (1981). Estimation of the mean of a multivariate normal distribution. *Ann. Statist.* **9** 1135–1151. MR0630098

STERN, H. (2005). Baseball decision making by the numbers. In *Statistics*: *A Guide to the Unknown*, 4th ed. (R. Peck, G. Casella, G. Cobb, R. Hoerl, D. Nolan, R. Starbuck and H. Stern, eds.) 393–406. Duxbury Press, Pacific Grove, CA.

STRAWDERMAN, W. E. (1971). Proper Bayes estimators of the multivariate normal mean. *Ann. Math. Statist.* **42** 385–388. MR0397939

STRAWDERMAN, W. E. (1973). Proper Bayes minimax estimators of the multivariate normal mean for the case of common unknown variances. *Ann. Math. Statist.* **44** 1189–1194. MR0365806

DEPARTMENT OF STATISTICS
UNIVERSITY OF PENNSYLVANIA
PHILADELPHIA, PENNSYLVANIA 19104
USA
E-MAIL: lbrown@wharton.upenn.edu